\documentclass[9pt,a4paper,twoside]{tau}
\usepackage[english]{babel}

\usepackage{graphicx}
\usepackage{amsmath,amssymb,amsfonts}
\usepackage{amsthm}
\usepackage{mathrsfs}
\usepackage{xcolor}
\usepackage{textcomp}
\usepackage{booktabs}
\usepackage{multirow}
\usepackage[table]{xcolor}
\usepackage{colortbl}
\usepackage[version=4]{mhchem}
\usepackage{siunitx}
\journalname{}
\title{Performance Evaluation of a Position-Sensitive SiPM-based Gamma Camera for Intraoperative Imaging}

\author[a]{Aramis Raiola}
\author[b]{Fabio Acerbi}
\author[a]{Cyril Alispach}
\author[a]{Domenico della Volpe}
\author[c]{Hossein Arabi}
\author[b]{Alberto Gola}
\author[c,d,e,f]{Habib Zaidi}

\affil[a]{Département de Physique Nucléaire et Corpusculaire (DPNC), Université de Genève, Geneva, Switzerland}
\affil[b]{Center for Sensor and Devices (SD), Fondazione Bruno Kessler (FBK), Trento, Italy}
\affil[c]{Division of Nuclear Medicine and Molecular Imaging, Geneva University Hospital (HUG), Geneva, Switzerland}

\affil[d]{Department of Nuclear Medicine and Molecular Imaging, University of Groningen, University Medical Center Groningen, Groningen, The Netherlands}

\affil[e]{Department of Nuclear Medicine, University of Southern Denmark, Odense, Denmark}

\affil[f]{University Research and Innovation Center, Óbuda University, Budapest, Hungary}

\institution{Université de Genève}
\footinfo{POSiCS / ATTRACT}
\theday{January 2026}
\leadauthor{A. Raiola et al.}
\course{POSiCS performance paper}
\begin{abstract}
\textbf{Background:} The POSiCS camera is a handheld and wireless, small field-of-view gamma camera developed for use in radio-guided surgery (RGS), with sentinel lymph node biopsy (SLNB) as its benchmark application. This compact and lightweight detector (about $350\,\mathrm{g}$) maps tissues labeled with a $\ce{^{99m}Tc}$ radiotracer to highlight target lesions. By enabling intraoperative visualization close to the surgical field, POSiCS aims to reduce invasiveness and operative time, improve localization accuracy, and minimize post-operative complications.

\medskip
\textbf{Methods:}
The performances of two hardware configurations were tested: high-sensitivity and high-resolution, to meet application-specific needs. Performance characterization employed Flood (flat) $\ce{^{99m}Tc}$ sources and $\ce{^{57}Co}$ point sources. The response to $\ce{^{177}Lu}$ was also investigated, given its growing role in theranostics. Tests assessed key parameters, including extrinsic spatial resolution, energy resolution, and sensitivity, as well as shielding effectiveness and response to different gamma energies. 

\medskip
\textbf{Results:} 
In direct contact, the spatial resolution was $1.9\pm0.1\,\mathrm{mm}$ in high-sensitivity mode and $1.4\pm0.1\,\mathrm{mm}$ in high-resolution mode. System sensitivity at $2\,\mathrm{cm}$ source-detector distance was $481\pm14\,\mathrm{cps/MBq}$ and $134\pm8\,\mathrm{cps/MBq}$, respectively. Both modes showed an energy resolution of about 20\% at 140 keV, though the high-resolution collimator exhibited increased scatter due to its larger tungsten content.

\medskip
\textbf{Conclusion:} 
The POSiCS camera combines compactness, ease of use, and reliable performance, making it suitable for RGS. Its interchangeable configurations allow balancing between sensitivity and resolution depending on surgical needs. The results demonstrate millimeter-scale resolution and adequate sensitivity for $\ce{^{99m}Tc}$ and $\ce{^{177}Lu}$ imaging, confirming its suitability for intraoperative use. These outcomes support the camera’s potential for broader clinical deployment and further development toward advanced intraoperative nuclear imaging.
\end{abstract}


\begin{document}

\maketitle
\tauabstract
\tableofcontents

\section{Introduction}\label{sec1}

Radio-guided surgery (RGS), including intraoperative techniques, such as sentinel lymph node biopsy (SLNB) and radioguided occult lesion localization (ROLL), is a widely adopted approach, particularly in the surgical management of breast cancer and cutaneous melanoma.~\cite{RGS-breast-cancer, Mariani811}. 
Despite the availability of well-established alternatives to RGS (such as 
Fluorescence-Guided Surgery (FGS) or Ultrasound-Guided Surgery), the preoperative administration of radio-labeled molecules remains one of the most widely used techniques for intraoperative localization of oncological lesions. 
This is due to the large mean free path of gamma rays in living tissues, which allows for the direct detection and localization of activity hot-spots from outside the patient's body~\cite{RGS-breast-cancer, Mariani811, Arabi_2024}.

\par The state-of-the-art practice in SLNB and ROLL is based on radiation-counting probes to localize lesions or sentinel lymph nodes (SLNs)~\cite{Tsuchimochi_2013, Farnworth2023}. Although this approach has been proven reliable by clinically comparing preoperative lymphoscintigraphies with the number of SLNs localized through a probe-based search, probes still inherently bear a number of drawbacks. 
To address this shortcoming, an increasing number of 2D Intraoperative Gamma Cameras (IGCs), also known as Small Field-of-View (SFOV) gamma cameras, have been developed in the last two decades~\cite{Stoffels2012, Massari2018}. 
This trend has been thoroughly studied in recent years, thanks to the comprehensive reviews in the field of IGCs, such as~\cite{Tsuchimochi_2013} and a more recent follow-up~\cite{Farnworth2023}. 
The main advantages of these cameras are their imaging capabilities, which allow surgeons to localize hot spots near the "blast zone" (radiotracer injection site) and to provide a method to verify post-excision images to ensure that the target lymph node or lesion was entirely resected~\cite{Hellingman_2016, Paredes2008}. Additionally, imaging is intuitive and does not require the specific and lengthy training needed to become proficient with probes.
It has also been shown that 2D imaging of oncological lesions can reduce surgery duration, as it facilitates the localization of target tissues more quickly~\cite{Hellingman_2016, soluri2006}. 

\par POSiCS (Position-sensitive SiPM Compact and Scalable gamma-Camera) is a handheld, compact, and wireless gamma camera designed for real-time intraoperative guidance. The device is based on a pixelated LYSO:Ce scintillator and a 3$\times$3 array of position-sensitive SiPMs (LG-SiPMs). This module is equipped with two interchangeable collimators: a short septa, low-energy, high-sensitivity (LEHS) collimator and a longer, low-energy, high-resolution (LEHR) collimator, enabling dual operating modes that can be selected intraoperatively to adapt to different clinical needs. Designed to effectively detect photons at $140.5\,\mathrm{keV}$, corresponding to the emission energy of $^{99\mathrm{m}}\mathrm{Tc}$~\cite{KleinJan2013}, the standard radionuclide used in RGS, the POSiCS camera exhibits enhanced performance relative to existing systems.

\par In this work, we present the results of the first performance studies on the prototype POSiCS camera module. In Section~\ref{ch:3}, we present the experimental methods adopted to measure the main performance parameters of the camera with both collimators. 
The tests were inspired from the 2023 NEMA standards for gamma camera qualification~\cite{NEMA2023}. However, as these standards are primarily designed for large clinical scanners, only a subset of the proposed measurements and procedures applies to our device, given its specific design and use case.
In Section~\ref{ch:4}, the results of the tests are displayed and commented. Conclusively, in Section~\ref{ch:5}, we provide an overall discussion of the obtained results, while in Section~\ref{ch:6}, we offer a brief outlook and prospects.

\section{Materials and Methods}
\label{ch:3}

\subsection{The POSiCS camera}
\label{ssec:corr}

\begin{figure}[t!]
    \centering
\includegraphics[width=\linewidth]{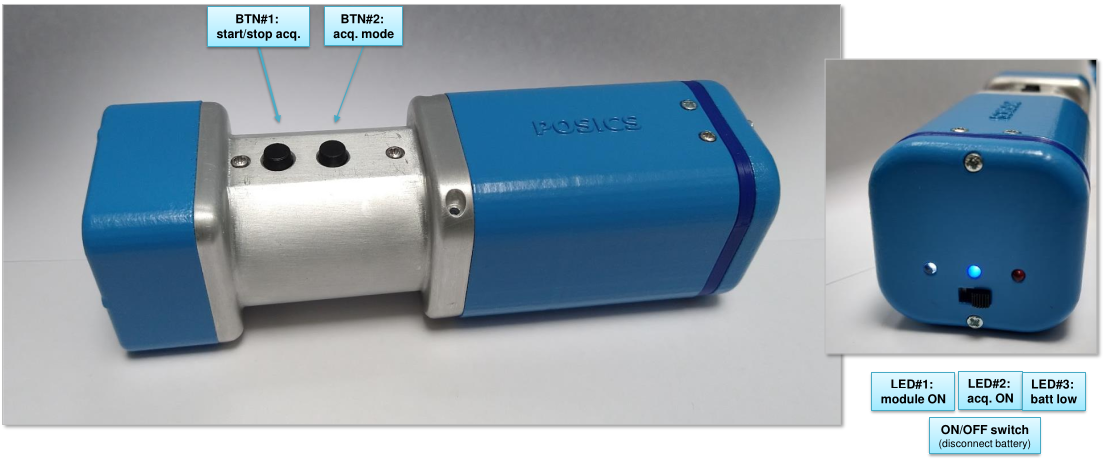}
    \caption{Prototype wireless module of the POSiCS gamma camera.}
    \label{fig:POSiCS_module}
\end{figure}

The current POSiCS camera module has a total length of 147 mm (155 mm with the LEHR collimator mounted), with a Field-of-View (FOV) of $30.8\times30.8~\rm{mm^2}$. From this, we define a Useful FOV (UFOV) of $28.6\times28.6~\rm{mm^2}$.
The module total weight is 313 g (with the LEHS collimator) and 381 g (with the LEHR collimator). Further technical details about the camera are provided elsewhere~\cite{Acerbi2025}. A picture of the camera's prototype wireless module is shown in~\autoref{fig:POSiCS_module}.

\subsubsection{Collimator geometry}
The POSiCS camera is equipped with two interchangeable parallel-hole collimators: one targeting high spatial resolution (LEHR) for surgical procedures that require precise identification of target lesion margins, and the other designed to achieve high sensitivity (LEHS) for faster scanning, even at low injected activities. The different parameters that qualify both objects were chosen through theoretical calculations and simulations using GATE~\cite{GATE2004}.
Due to its high atomic number (implying a short attenuation length) and good resistance to shocks~\cite{Farnworth2023}, tungsten was chosen. All collimator parameters were chosen to ensure a septal penetration probability of less than $5\%$~\cite{collimator-design, cherry_ch14}. The collimator was designed with external shielding walls that extend beyond the collimator holes and support the scintillator, reducing the amount of gamma rays entering from the side of the camera. 
\begin{figure*}
    \centering
     \includegraphics[height=0.54\textheight, angle=90]{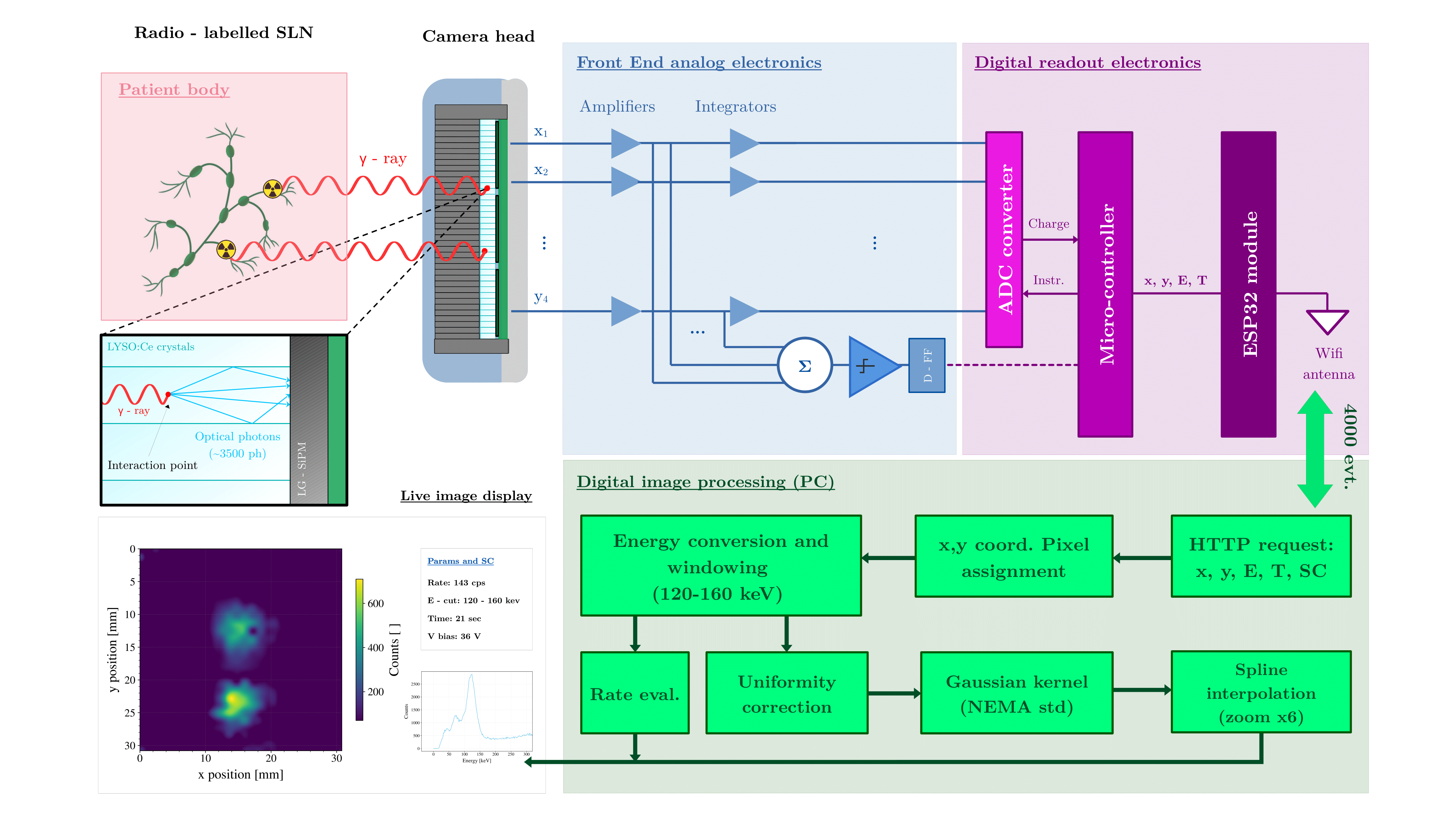}
    \caption{Scheme of the full pipeline of the POSiCS camera, from the gamma-ray emission by radio-labelled tissues, to the final image treatment and enhancement, with the possibility of live displaying the reconstructed image.}
    \label{fig:POSiCS_operation}
\end{figure*}

\subsubsection{LYSO scintillator approach}
\label{ssec:scintillator}
The radiation-sensitive component of the POSiCS camera is a pixelated LYSO:Ce scintillator. This Lutetium-based crystal is typically used in Positron Emission Tomography (PET) scanners due to its fast decay time ($\sim 40~\rm{ns}$), high light yield ($\geq32000~\rm{ph~MeV^{-1}}$), and high density~\cite{Kimble_2002}. The camera was therefore built using a very thin crystal (a few millimeters thick), keeping a high photoelectric absorption probability at $140.5~\rm{keV}$. The scintillator is pixelated in a 30$\times$30 matrix. Each pixel is coated with a BaSO$_4$ reflector to increase light collection on the light detector.
\par A peculiarity of LYSO:Ce is its intrinsic radioactivity due to the natural 2.6\% abundance of $^{176}$Lu isotope, which decays into $^{176}$Hf through a $\beta^{-}$ decay. The process is followed by the emission of three gamma rays at energies: $88~\rm{keV}$, $202~\rm{keV}$, and $307~\rm{keV}$, which can interact within the crystal itself~\cite{Conti_2017, LYSO-Lutetium176}. The presence of this isotope establishes a background activity within the camera FOV at about $420~\rm{cps}$, which is mainly eliminated through energy window selection and subtracted during image correction. This feature is often cited as being suited for self-calibration in PET scanners and can represent an advantage in this direction in the future~\cite{Wei_2014}. The LYSO:Ce intrinsic spectrum reconstructed with the POSiCS camera is displayed in~\autoref{fig:flood_setup}.

\subsubsection{LG-SiPMs photodetector}
Until 2013, reviews in the field of IGCs showed a trend toward implementing Position-Sensitive (PS-) PMTs in SFOV gamma cameras~\cite{Tsuchimochi_2013}. At the same time, most recent reviews highlight a shift in the past decade toward using SiPMs for the optical readout of scintillator crystals in IGCs~\cite{Farnworth2023}. This transition is driven by the advantages of solid-state detectors, including low power consumption, compactness, high quantum efficiency, and insensitivity to magnetic fields.
\par In the current context of light detectors for IGCs, POSiCS proposes an innovative approach by moving towards PS-SiPMs, leveraging the new Linearly Graded SiPMs technology (LG-SiPMs) designed and produced by FBK~\cite{Acerbi2024,FBK-LG-SiPM,Acerbi2024_ps}. Through an array of 3x3 LG-SiPMs with peak sensitivity at $420~\rm{nm}$, an active area of about $10\times10~\rm{mm^2}$ per sensor, and microcell pitch of $25~\rm{\mu m}$, it is possible to reconstruct the position of each scintillation event, through a weighted charge (and amplitude) distribution towards the output channels. Each Single Photon Avalanche Diode (SPAD) composing the SiPM is connected to a resistive network, which performs a light spot CoG reconstruction, encoded in the relative amount of charge output by each readout channel~\cite{Acerbi2025, Acerbi2024}. This Anger CoG reconstruction is typically used between SiPMs or PMTs in standard gamma cameras, but this approach is implemented at the single microcell level within the POSiCS camera. The 3$\times$3 array configuration, with the "smart channel" approach (see~\cite{Acerbi2024}), allows CoG reconstruction with only 8 readout channels: 4 encoding for the x coordinate ($x_1,x_2,x_3,x_4$) and 4 reconstructing the y coordinate ($y_1,y_2,y_3,y_4$).  More precisely, position reconstruction is performed as follows:
\begin{subequations}
\begin{align}
    x_{pos} &= \frac{x_1+\frac{1}{3}(x_3-x_2)-x_4}{x_1+x_2+x_3+x_4}, \\
    y_{pos} &= \frac{y_1+\frac{1}{3}(y_3-y_2)-y_4}{y_1+y_2+y_3+y_4}.
\end{align}
\label{eq:LG_SiPM}
\end{subequations}
This weighted position calculation is performed directly within the camera digital electronics, which in turn sends the reconstructed positions through a wireless connection.
This photon detector demonstrated exceptional intrinsic precision, achieving a resolution of a few hundred microns~\cite{Acerbi2024, Gola2013ANA, raiola2025spatialresolution}. 
\par The limited number of channels enables a fast and straightforward readout, allowing for wireless data transmission to the computer before applying the necessary image quality processing.

\subsubsection{Approach to evemt reconstruction}

After a scintillation event in the scintillator, the eight SiPM signals are amplified by transimpedance amplifiers (TIAs). Their summed charge output is sent to a comparator, which enables charge integration only when a user-defined threshold is crossed. The resulting per-channel charge is integrated and digitized by an eight-channel ADC. The microcontroller (\(\mu\)C) then computes the event energy \(E\) (sum of all channels output charge), timestamp \(T\), and the \(x\)- and \(y\)-coordinates using~\autoref{eq:LG_SiPM}, and also provides slow-control (SC) data such as bias voltage, current, temperature, and battery status (see~\autoref{fig:POSiCS_operation}).

Each event is stored as \((x, y, E, T)\) and transmitted to an external computer via an ESP32 module in packets of 4000 events or after a user-defined time window has elapsed. The data is retrieved via HTTP GET requests. The camera settings are configured through the same ESP32 using HTTP POST commands to the DACs.
For further technical details about the POSiCS camera electronics, see~\cite{Acerbi2025}.

Reconstructed images are corrected using a three-step calibration and correction procedure. First, once the incident position of a gamma ray is determined (x,y CoG coordinates), the event is assigned to the corresponding camera pixel (scintillator crystal) associated with the scintillator crystal encompassing the interaction position. This spatial mapping is established through a flood-field calibration scan, which serves as the primary calibration step before operating the camera.
\par Second, a pixel-by-pixel energy calibration is performed. In this step, a scaling coefficient is applied to convert the SiPMs charge response into a physical energy. Due to non-uniform light collection across certain regions of the FOV, as well as slight variations in the gain and response characteristics of the SiPMs forming the optical readout array, each pixel is assigned an individual conversion coefficient to convert the digitized charge to a physical energy in keV. Following this step, a user-defined energy window can be applied in software to selectively retain events consistent with photoelectric interactions, thereby suppressing contributions from Compton-scattered events that occur either within the LYSO crystal or in the surrounding environment.
\par Finally, a uniformity correction is applied to the reconstructed image to compensate for spatial sensitivity variations. This step enhances the signal in regions typically exhibiting under-response, such as those corresponding to interstitial gaps between SiPMs or at the periphery of the camera field of view, thereby ensuring a more homogeneous image representation.
\par With this acquisition system, it is possible to display images live with a delay of the order of 1 second. The final images are enhanced using a 2D-Gaussian kernel convolution, following the guidelines in the 2023 NEMA standards for the validation of gamma cameras~\cite{NEMA2023}. Finally, the smoothed image is interpolated using splines of order 2.

\begin{figure*}[t!]
  
    \includegraphics[width=\linewidth]{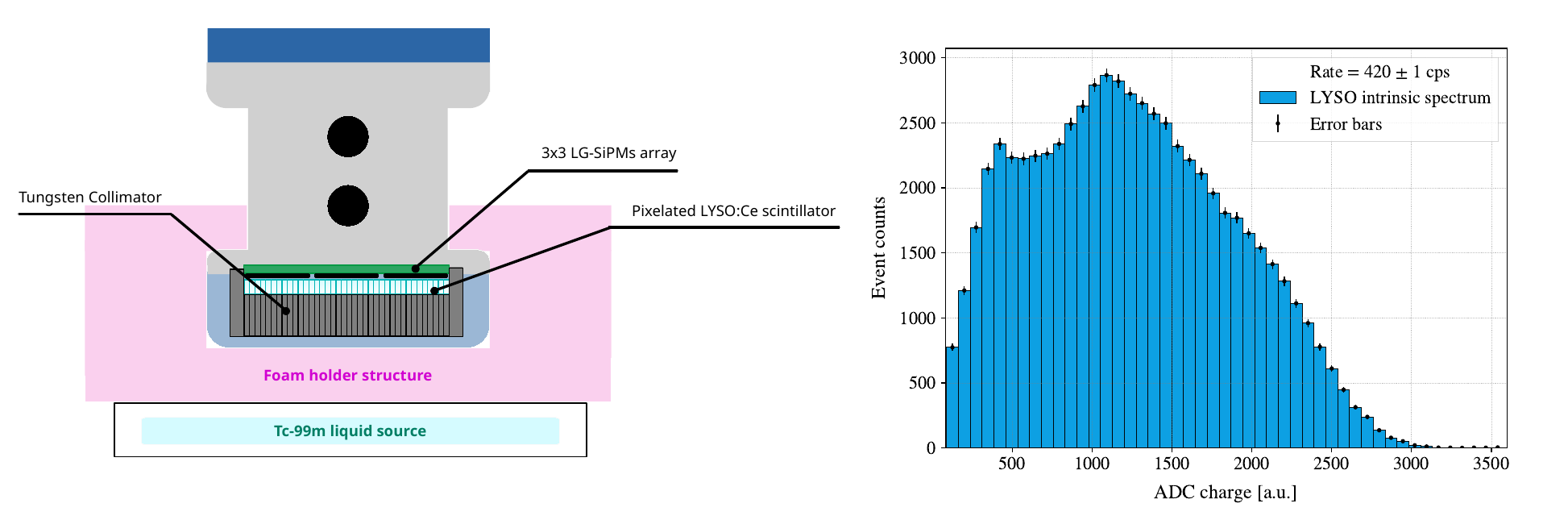}
    \caption{\textbf{Left: }Scheme of the flood map setup used in this work. The estimated distance from the collimator entrance to the center of the cavity is 2 cm. \textbf{Right: }LYSO:Ce scintillator intrinsic radioactivity spectrum registered with POSiCS. The structure of the spectrum depends on the crystal volume and shape, which in turn determines the probability of absorbing the emitted $\gamma$-rays (in coincidence). The long tails are due to the absorption of the initial $\beta$ particle and are therefore shaped after the $\beta$-emission spectrum~\cite{LYSO-Lutetium176}. Tc-99m photoepeak is expected to lie at 460 ADC counts.}

    \label{fig:flood_setup}
\end{figure*}

\subsection{Performance evaluation}

The POSiCS camera underwent extensive performance testing to evaluate the reliability of its image reconstruction algorithms and to characterize its applicability across different imaging scenarios. The key performance parameters investigated included spatial resolution, sensitivity, and energy resolution. These evaluations required the use of both point-like and extended radioactive sources.  

To assess spatial resolution, a point source, defined as having a diameter smaller than that of the scintillator pixel, was necessary. Due to the practical limitations of producing such a source with \(\mathrm{^{99m}Tc}\), a solid \(\mathrm{^{57}Co}\) source was employed. This isotope emits gamma-rays at 122~keV, serving as a suitable surrogate for the 140.5~keV photons of \(\mathrm{^{99m}Tc}\).  

Conversely, the extrinsic sensitivity and energy resolution tests with \(\mathrm{^{99m}Tc}\) were performed using a custom-designed flat liquid phantom that provided uniform irradiation of the detector, as illustrated in~\autoref{fig:flood_setup}. Finally, additional tests were conducted with \(\mathrm{^{177}Lu}\), a radionuclide used in theragnostic that emits multiple photon energies ranging from low-energy X-rays to \(\gamma\) emissions above 300~keV. These measurements aimed to evaluate the camera’s reconstruction performance with multi-energy emissions and to determine its suitability for imaging with \(\mathrm{^{177}Lu}\).

\begin{table*}[t]
\centering
\renewcommand{\arraystretch}{1.3}
\begin{tabular}{lccc>{\centering\arraybackslash}p{3cm}}
\hline\hline
\textcolor{blue}{\textbf{Isotope}} & 
\textcolor{blue}{\textbf{Energy [keV]}} & 
\textcolor{blue}{\textbf{Window [keV]}} &
\textcolor{blue}{\textbf{Source form}} & 
\textcolor{blue}{\textbf{Test}} \\
\hline\hline

\(\mathrm{^{57}Co}\)   & 122     & 100--140 & Sealed point source & Spatial resolution, collimator leakage \\
\(\mathrm{^{99m}Tc}\)  & 140.5   & 120--160 & Flat liquid phantom & Sensitivity, Energy resolution \\
\(\mathrm{^{177}Lu}\)  & 55--208 & Multiple & Liquid source & Multi-energy reconstruction \\

\hline\hline
\end{tabular}

\caption{Summary of the radioactive sources and test conditions used to evaluate the POSiCS camera performance.}
\label{tab:source_tests}
\end{table*}

\subsubsection{Extrinsic Spatial resolution}
\begin{figure}[h!]
    \centering
    \includegraphics[width=0.8\linewidth]{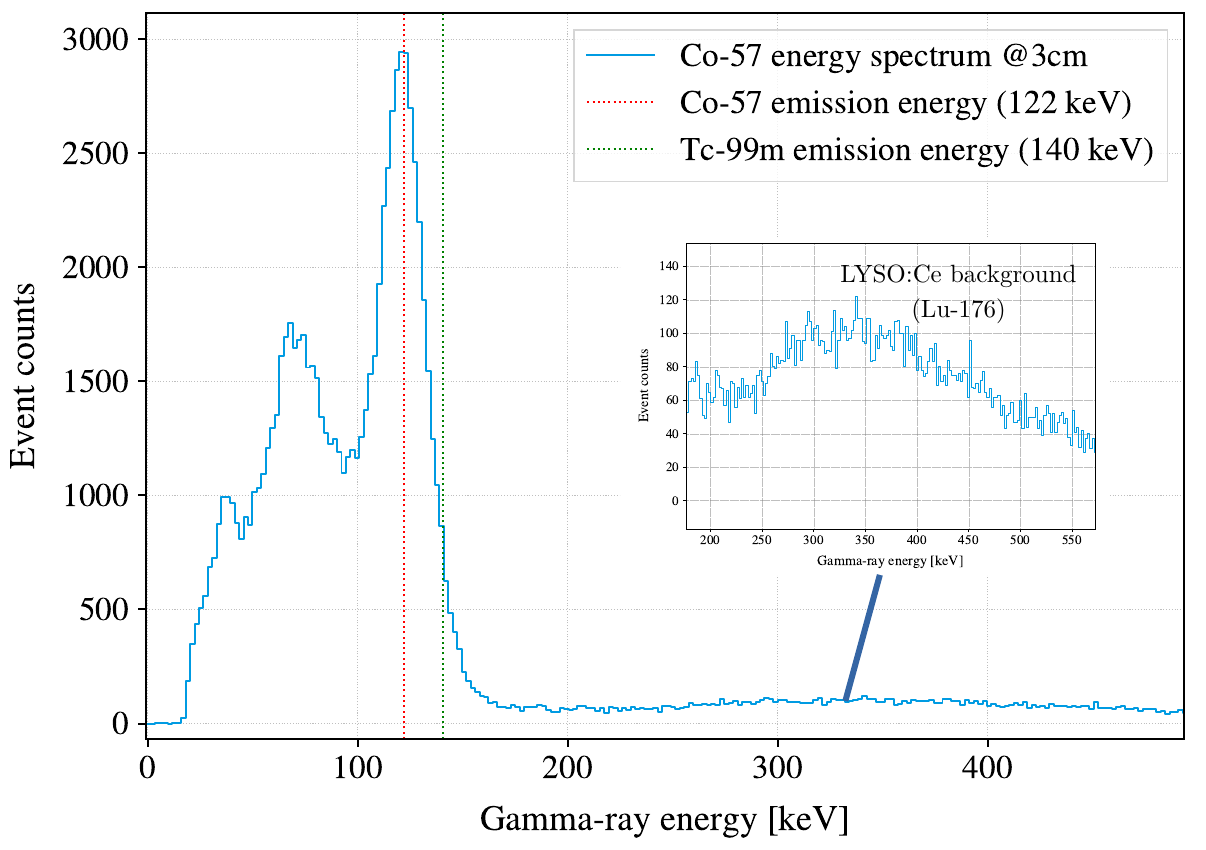}
    \caption{Example of energy spectrum of the Co-57 point source positioned 3 cm from the center of the camera FOV (LEHR collimator). The background-dominated region associated with LYSO intrinsic radioactivity is highlighted, illustrating that it does not significantly contribute to the energy window centered around 122 keV. For reference, the position of the Tc-99m emission line at 140.5~keV is also reported.}
    \label{fig:Co57_spectrum}
\end{figure}

The extrinsic spatial resolution was assessed by placing a point-like $^{57}$Co source ($E_\gamma=$122 keV) in the center of the camera's FOV. In this position, 100'000 events were acquired, and an image was reconstructed following the correction pipeline described in subsection~\ref{ssec:corr}. No smoothing or kernel convolution is applied to assess the device's performance without digital image quality enhancement. An energy cut window between 100 and 140 keV was applied to reject scattered photons (cf. Figure~\ref{fig:Co57_spectrum}). The obtained reconstructed $x$ and $y$ profiles were fitted with a Gaussian profile, and their full width at half maximum (FWHM) was calculated. We define the device's resolution as the average FWHM between the two directions.

\par The uncertainty on the resolution is the summation in quadrature of the statistical error (obtained through the covariance matrix of the fit) and a systematic uncertainty. Systematics is defined as the variation in resolution due to the different behavior of the camera in other areas of the FOV. The systematic uncertainty is evaluated through a scan of 15 positions on the device's FOV with the $^{57}$Co point source, acquiring 100'000 events at each position. Each reconstructed spot is fitted with a Gaussian profile, from which the FWHM is extracted. The standard deviation of this sample of 15 measured resolution values represents the systematic error of the resolution measurement.

\par The measurement at the center of the FOV is repeated for different source distances and under two conditions: with and without scattering material between the source and the camera head. The chosen scattering material is Polymethyl methacrylate (PMMA). This approach enables us to estimate the expected resolution in biological tissues, whose density is comparable to that of PMMA. The obtained resolution is also compared with the theoretical spatial resolution~\cite{cherry_ch14,AngerCamera, xcom2010}:

\subsubsection{Sensitivity}
The camera's sensitivity is defined as the registered trigger rate divided by the activity of the source within the camera's FOV. This parameter is strongly dependent on the geometry of the collimator, the efficiency of the scintillator crystal (i.e., the probability of absorbing a gamma ray through photoelectric effect at a defined energy), and the dead-time of the readout. Following the guidelines presented in the NEMA standards for gamma camera qualifications~\cite{NEMA2023}, we performed sensitivity measurements, ensuring that rates did not exceed 20'000 counts per second (cps). 
\par The tests were performed by placing the flood phantom filled with Tc-99m dissolved in saline solution, to irradiate the FOV uniformly. Then, the rate was registered through the camera's wireless readout, in packets of 4000 events. The activity of the source visible in the FOV was evaluated through the following conversion:
\begin{equation}
    A_{\rm{FOV}}(t)=A(t_0)\cdot  \frac{S_{\rm{FOV}}}{S_{\rm{FP}}} \cdot e^{-\frac{t\ln{2}}{\tau_{1/2}}}
\end{equation}
where $t_0$ is the source's production time (when activity $A(t_0)$ was measured by a $4\pi$ detector available at Geneva University Hospital, with a precision down to tens of kBq), $\tau_{1/2}=6.02~\rm{h}$ is the half-life of metastable Tc-99m, $S_{\rm{FOV}}$ is the area of the FOV, and $S_{\rm{FP}}$ is the surface of the Flood Phantom. The sensitivity over time is therefore computed as:
\begin{equation}
    g(t)=\frac{R(t)}{A_{\rm{FOV}}(t)}
\end{equation}

Here, $R(t)$ is the trigger rate of the camera at time $t$. The reported error on the sensitivity measurements is defined as the quadrature sum of the statistical uncertainty on the rate and the systematic uncertainty on the activity, giving the uncertainty in sensitivity. The latter is the propagated error of the initial injected activity in the phantom (set at $10\%$ of the initial activity) and the uncertainty on the custom-made phantom's dimension:
\begin{equation}
    \sigma_{sens}=\sqrt{\frac{R(t)}{\Delta t}}\oplus\sigma_{syst}
\end{equation}
The procedure is repeated with the LEHS collimator and the LEHR collimator, placing the source at 2cm from the camera's lid. No background subtraction was performed, due to the high activity of the source with respect to the $\sim$500 cps rate from LYSO:Ce, which is reduced to few tens of cps after energy window selection. 

\subsubsection{Energy resolution}

The energy resolution was evaluated for both the LEHR and LEHS collimators by placing the Flood phantom in front of the camera to uniformly irradiate the FOV at a distance of 2 cm from the source. The resulting spectrum was corrected using a pixel-by-pixel approach, following the device's standard calibration procedure. The resolution was computed by fitting the photopeak (identified as the first and most prominent peak on the right, corresponding to the physical interaction process that leaves the most energy in the crystal) with a Gaussian profile and considering the resolution as its FWHM divided by the central theoretical energy of 140.5 keV.

\subsubsection{Collimator side penetration}
The camera's external shielding was tested to assess the penetration of gamma rays from the Tungsten collimator's walls and the septal penetration, which are designed to prevent external radiation from reaching the scintillator crystal. An IGC does not require strong directional sensitivity, unlike 1D gamma probes, which rely on it due to their lack of imaging capability. Directionality is, therefore, a key performance aspect typically studied in detail for probes~\cite{Wydra2005}. In the case of POSiCS, a penetration study was performed to ensure that possible activity hotspots outside the FOV do not influence the imaging performance of the POSiCS camera. Moreover, penetration of gamma-ray from the side are ultimately visually identified and thus can be removed by truncating the image.
\par This test was performed by registering the detected rate of a Co-57 point source in free air, positioned 3 cm from the center of the camera, at three distinct angles: one at $0^\circ$ relative to the normal of the FOV, and the other two at $45^\circ$ and $-45^\circ$ from the same normal direction. The 100-140 keV energy window (used throughout this work for Co-57 data processing) was adopted. 
\par The rate data was collected in 150 packets of 4000 events for each tested angle. For each packet, the initial and final timestamps are used to compute the average rate of that packet.

\subsubsection{Energy response to other radiotracers (Lu-177)}
To test the camera's performance at different $\gamma$-rays energies, we performed a study using Lu-177 as a radiotracer. This isotope has been adopted in nuclear medicine due to its multiple applications in radiotherapy~\cite{Dias2010, Ljungberg2016, Marin2017, Ramonaheng2024, Sagisaka2024}. Lu-177 is often linked to peptides (such as the case of Lu$^{177}$-DOTA-TATE and Lu$^{177}$-PSMA), to target cancerous cells in several radiotherapeutic procedures, aiming to treat prostate cancer, neuroendocrine tumors, bone metastasis, and other malignancies~\cite{Marin2017}. Lu-177 nuclei decay into Hf-177 through $\beta^{-}$ emission. Hf-177 nuclei live in a metastable state for characteristic times of a few hundred picoseconds (depending on the state), reaching stability through the emission of gamma rays~\cite{Dias2010}. The emission energies issued by the above-mentioned gamma dis-excitation: 208 keV (10.4\% probability), 113 keV (6.2\% probability), 55 keV (4.5\% probability), and 65 keV (1.2\% probability)~\cite{Marin2017, Sagisaka2024}. Some other, less frequent energies are also present. 

\par To perform dosimetry measurements on patients during therapy, SPECT scanners have already been used to map the activity regions traced by Lu-177 gamma-rays emission~\cite{Ljungberg2016}. Moreover, SFOV gamma cameras were also proposed and tested for their response to Lu177, due to the possibility of rapid scans with a smaller and more practical device~\cite{Roth2020}. The use of handheld cameras can also contribute to pharmacokinetic studies, allowing for faster and more frequent scans~\cite{Roth2024}.

\par The performance of the POSiCS camera was evaluated by imaging a~$204~\mathrm{mm}^3$ cavity, representing the rear tumor volume within a BIOMETECH fillable mouse phantom~\cite{BiometechMousePhantom}, depicted in~\autoref{fig:Lu177_images}. The cavity was filled with 12~MBq of Lu-177. The phantom was positioned in direct contact with the POSiCS camera, which was equipped with the LEHS collimator. Data were acquired over a total acquisition time of 100 seconds.

\section{Results}
\label{ch:4}
\subsection{Extrinsic spatial resolution}
\par
\begin{figure*}[t!]
    \centering
    \includegraphics[width=0.47\linewidth]{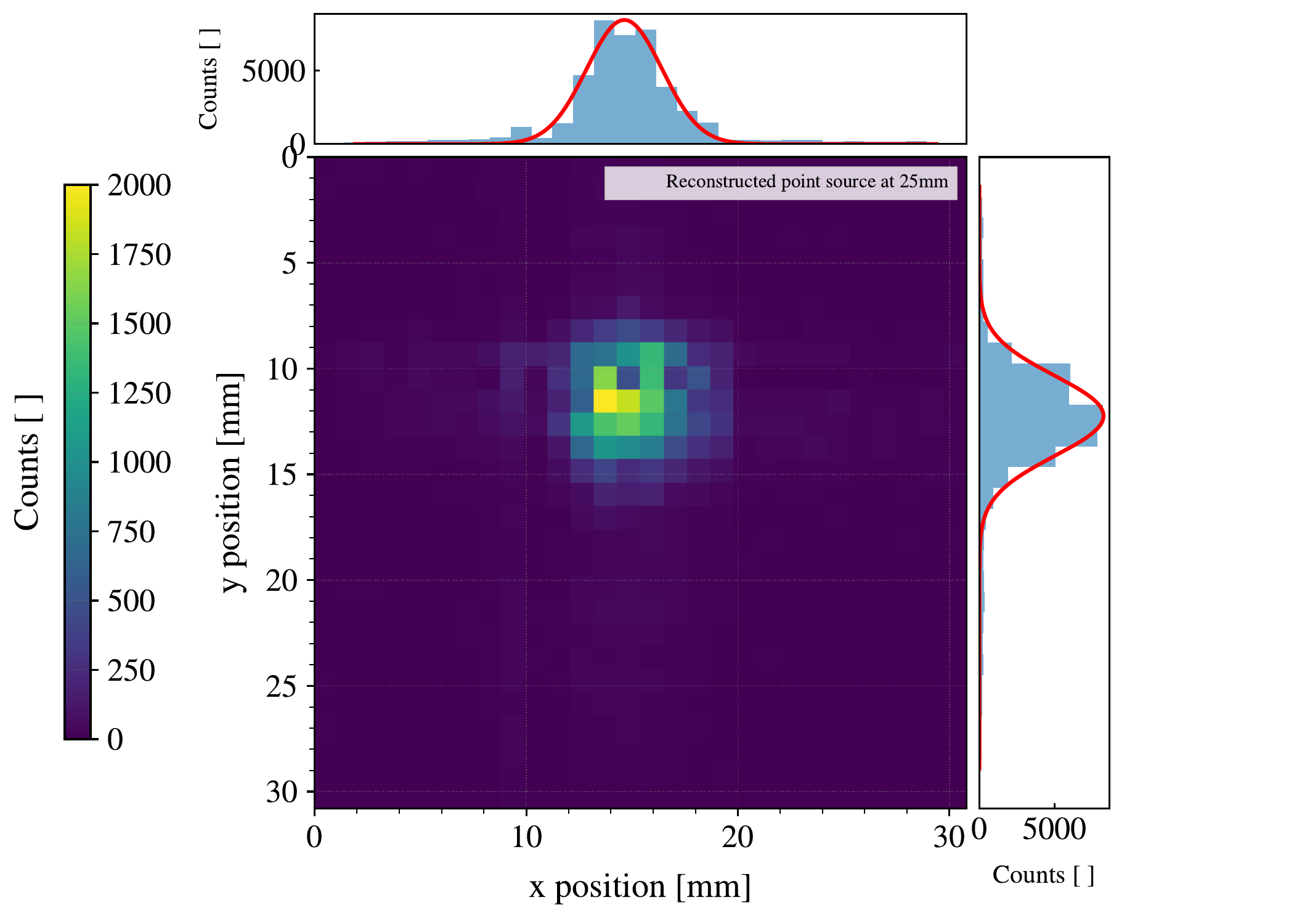}
    \includegraphics[width=0.52\linewidth]{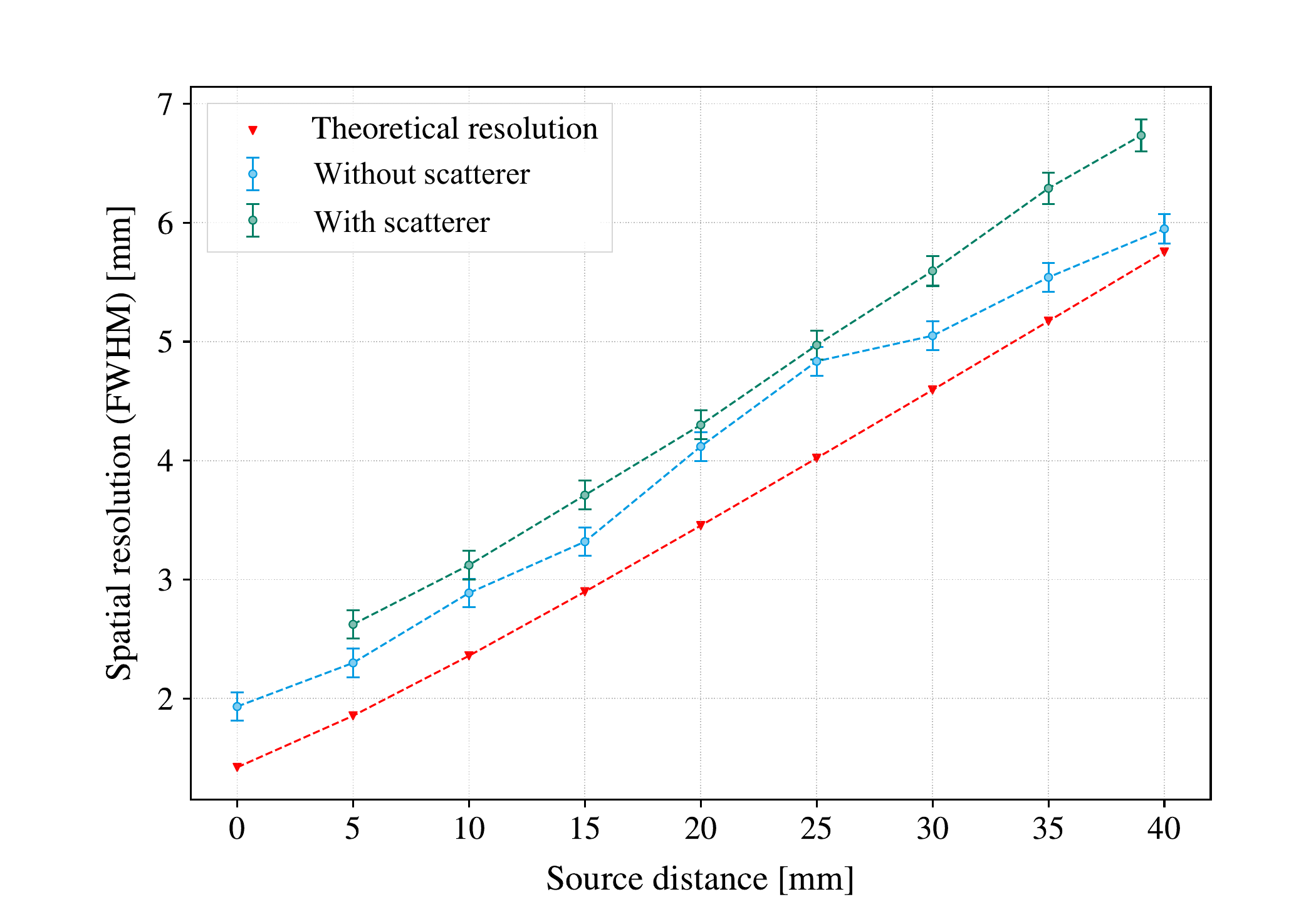}\\
     \includegraphics[width=0.47\linewidth]{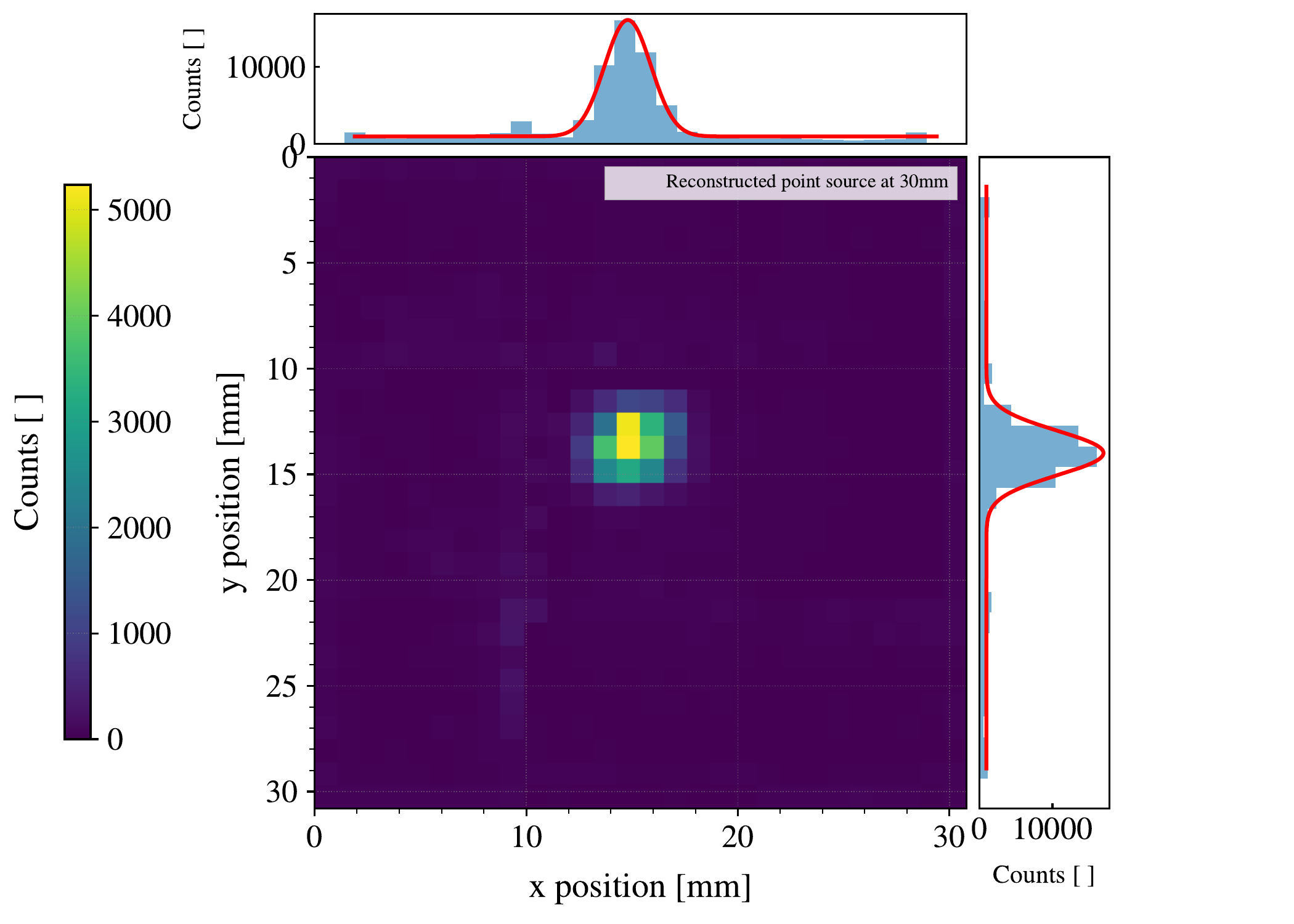}
    \includegraphics[width=0.52\linewidth]{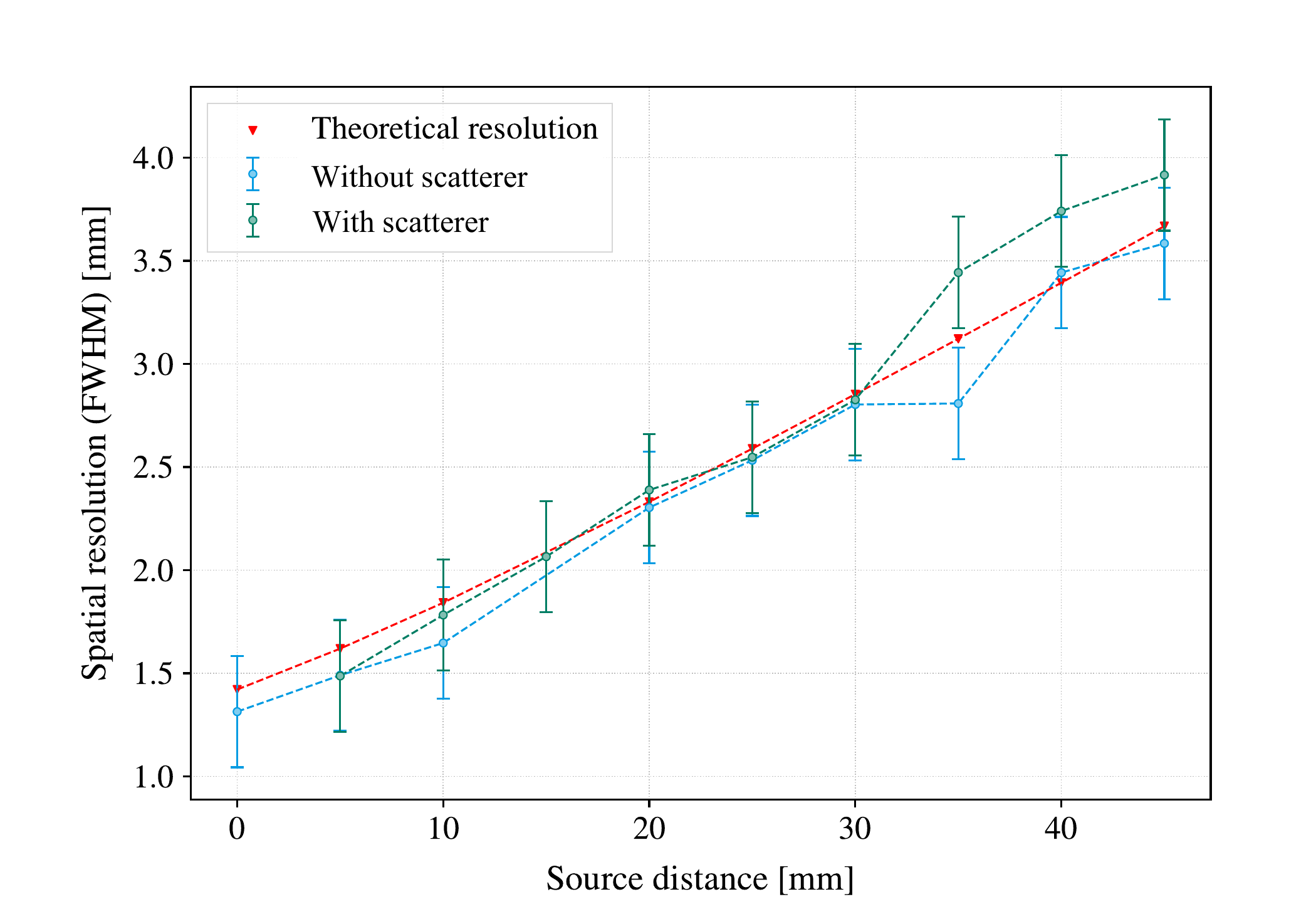}
    \caption{\textbf{Top Left:} Example of the reconstructed point source at 3~cm from the camera (free air) with LEHS collimator. \textbf{Top Right:} Resolution of the POSiCS camera in HS mode for different distances. \textbf{Bottom Left:} Example of the reconstructed point source at 3~cm from the camera (free air) with LEHR collimator. \textbf{Bottom Right:} Resolution of the POSiCS camera in HR mode.}

    \label{fig:resolution}
\end{figure*}

 The results of the spatial resolution evaluation are presented in Figure~\ref {fig:resolution}. 

By placing the source in contact with the camera, we obtained a resolution with the LEHR collimator of 1.4~$\pm$~0.1~mm, which increased to 2.8~$\pm$~0.1~mm when the source was moved 3~cm away. In contact with the camera head, the LEHR and LEHS collimators show comparable resolutions. However, due to the well-known collimator projective effect~\cite{cherry_ch14, AngerCamera}, the resolution of the LEHS collimator degrades faster than for the LEHR collimator. With the LEHR collimator, we measured a spatial resolution of 1.9~$\pm$~0.1~mm at 0~cm from the source, which increased to 5.1~$\pm$~0.1~mm at 3~cm. Moreover, the difference between measurements in free air and with scatterer material is within error bars using the LEHR collimator, while the presence of PMMA worsens resolution with the LEHS collimator. This phenomenon is likely related to the collimator's improved ability to reject photons scattered in the PMMA layers between the source and the camera. 
\par For the LEHS collimator, the measured extrinsic resolution in free air is larger than the theoretical one by 15\% on average, as displayed in~\autoref{fig:resolution} (top Right). This inconsistency, which lies beyond error bars, is mainly due to the light detector and the thin collimator septa. Theoretical calculations, indeed, do not consider the resolution spread due to the light detector, which, in our case, presents some deformations toward the interstitial gaps between SiPMs, for instance. Moreover, for a short collimator with thin septa, it is possible to have an increased septal penetration, and therefore a reduced spatial resolution, which is challenging to estimate empirically. Consequently, we can conclude that the camera's extrinsic resolution aligns reasonably well with theoretical expectations, showing a similar slope as the distance increases. The LEHR collimator analysis reveals a more precise agreement with theoretical expectations, likely due to the reduced number of accepted scattered photons and reduced septal penetration.

\subsection{Extrinsic sensitivity}
\begin{table}[ht]
\centering
\renewcommand{\arraystretch}{1.4}
\begin{tabular}{lcc}
\hline
  
\textbf{\textcolor{blue}{Sensitivity results}} & \textbf{\textcolor{blue}{LEHS coll.}} & \textbf{\textcolor{blue}{LEHR coll.}} \\
\hline\hline
Sensitivity BEC [cps/MBq] &1444 $\pm$ 31 &730 $\pm$ 16 \\
Sensitivity AEC [cps/MBq]&481 $\pm$ 14 &134 $\pm$ 8 \\
Distribution Kurtosis & 2.89 & 2.97 \\
Distribution skewness & 0.04 & 0.00 \\

\hline\hline
\end{tabular}
\caption{Sensitivity results compared to the theoretical expectations. BEC = Before Energy Cut, AEC = After Energy Cut.}
\label{tab:sensitivity}
\end{table}
For the LEHS mode camera, we report a sensitivity of $481\pm14~\rm{cps/MBq}$ after rejecting all events whose energy does not lie within the spectral window adopted in this work for Tc-99m (120 keV-160 keV). Instead, we observe a significant reduction in sensitivity for the LEHR collimator, down to $134\pm8~\rm{cps/MBq}$, as expected. 

\par
\begin{figure}[t!]
    \centering
    \includegraphics[width=1\linewidth]{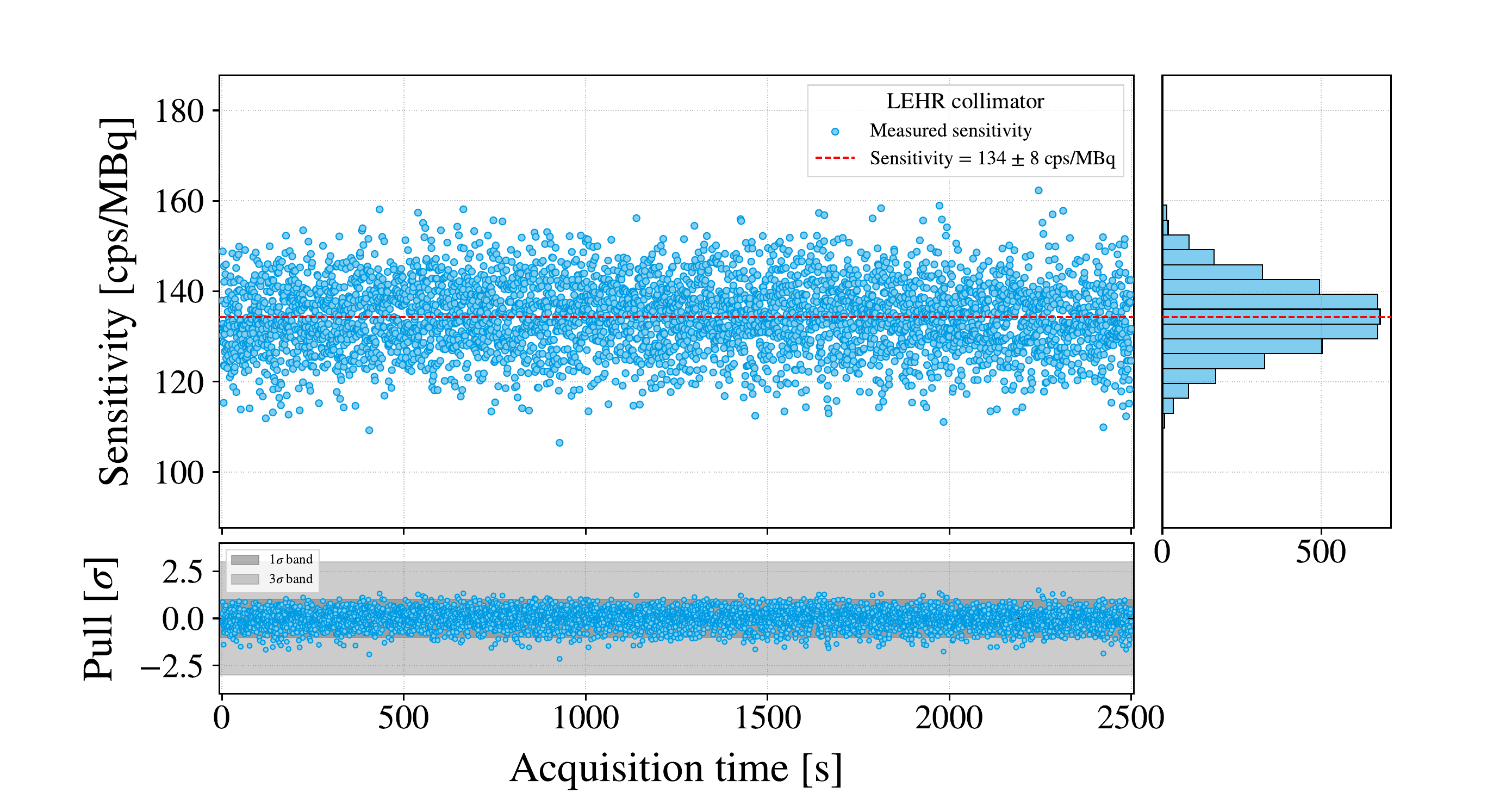}\\
    \includegraphics[width=1\linewidth]{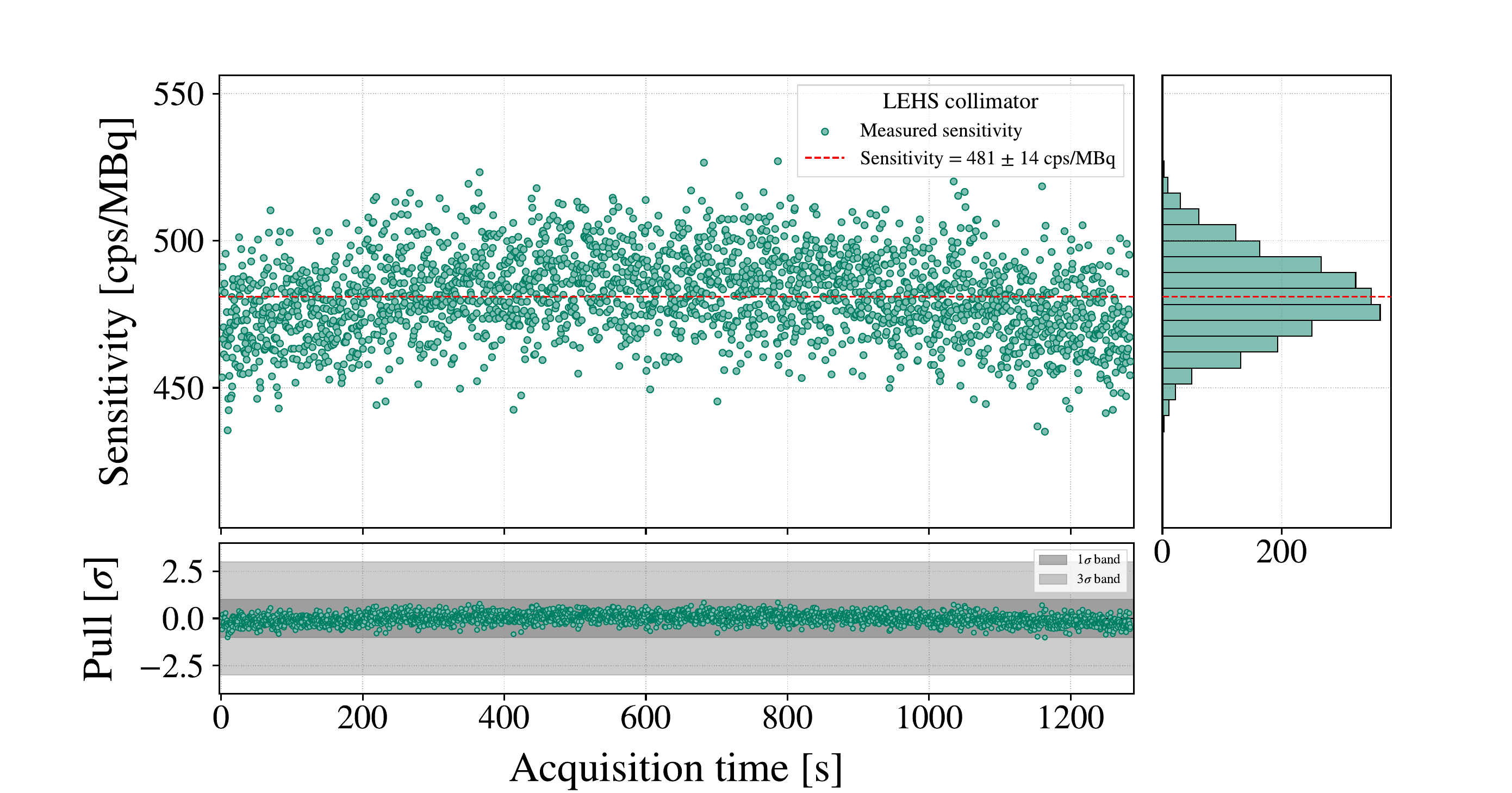}
    \caption{\textbf{Top:} Sensitivity over time of the camera in HR mode at 2cm from the source, after energy window cut (120-160 keV). \textbf{Bottom:} Sensitivity over time of the camera in HS mode at 2cm from the source, after energy window cut.}
    \label{fig:sensitivity}
\end{figure}

 The results of the sensitivity evaluation are displayed in~\autoref{fig:sensitivity} for both collimators. For the camera in HR mode, we observed a measured sensitivity distributed as a Gaussian function around its central value, with a standard deviation of 8 cps/MBq. The mean and the standard deviation were weighted by the points variance. This is displayed in the sensitivity projection on the right side-plot. The Gaussianity of this distribution implies that the negative exponential decay correction to the activity matches the observed rate. This suggests that the fluctuations after the source's activity compensation are mostly driven by poissonian shot-noise. Furthermore, from~\autoref{tab:sensitivity}, we evince that the Skewness and Kurtosis of the points distribution (side plot) are compatible with a Gaussian distribution, for which the typical skewness is 0 and the typical kurtosis is 3. For the LEHS collimator, we observe a distribution of sensitivities characterized by a slight positive skewness (indicating a heavier right tail) and a kurtosis slightly below 3, which suggests a more peaked distribution. This is also evident in the time series of measured sensitivities, which shows a positive vertical bending of the estimated sensitivity around 600 s (cf.~\autoref{fig:sensitivity}). Since the kurtosis and skewness analyses still suggest a mostly Gaussian behaviour, we cannot explain this visual bending at this stage as a systematic feature. On the other hand, it is possible that saturation effects due to the high count rate introduce systematic distortions in the time series. This feature will be further investigated. Due to the possibility of saturation, especially at the beginning of the acquisition time (when the source activity is maximal), we consider the measured extrinsic resolution of the LEHS collimator as a lower limit. 
\subsection{Energy resolution}

\begin{figure*}[t!]
    \centering
    \includegraphics[width=0.49\linewidth]{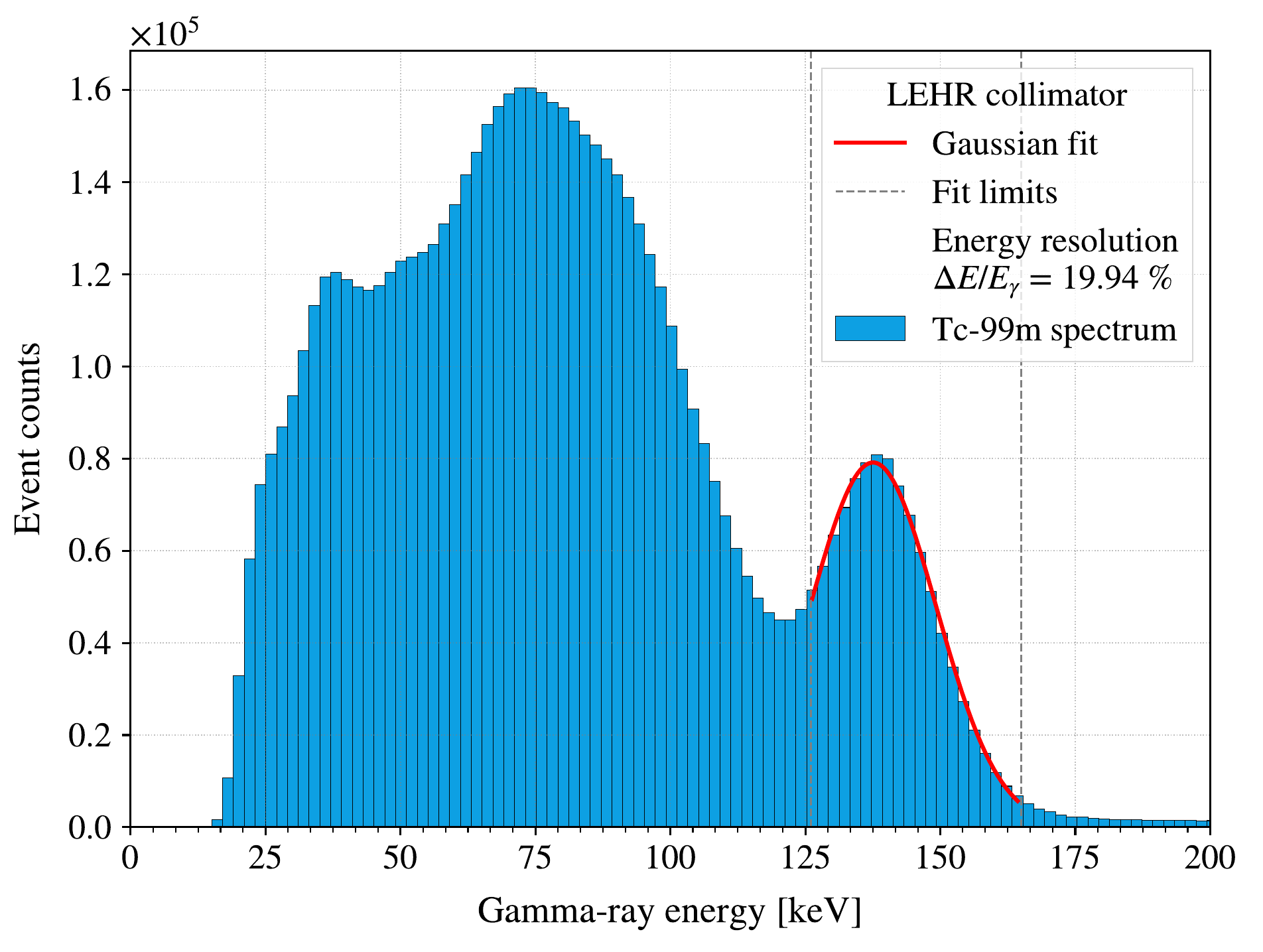}
    \includegraphics[width=0.49\linewidth]{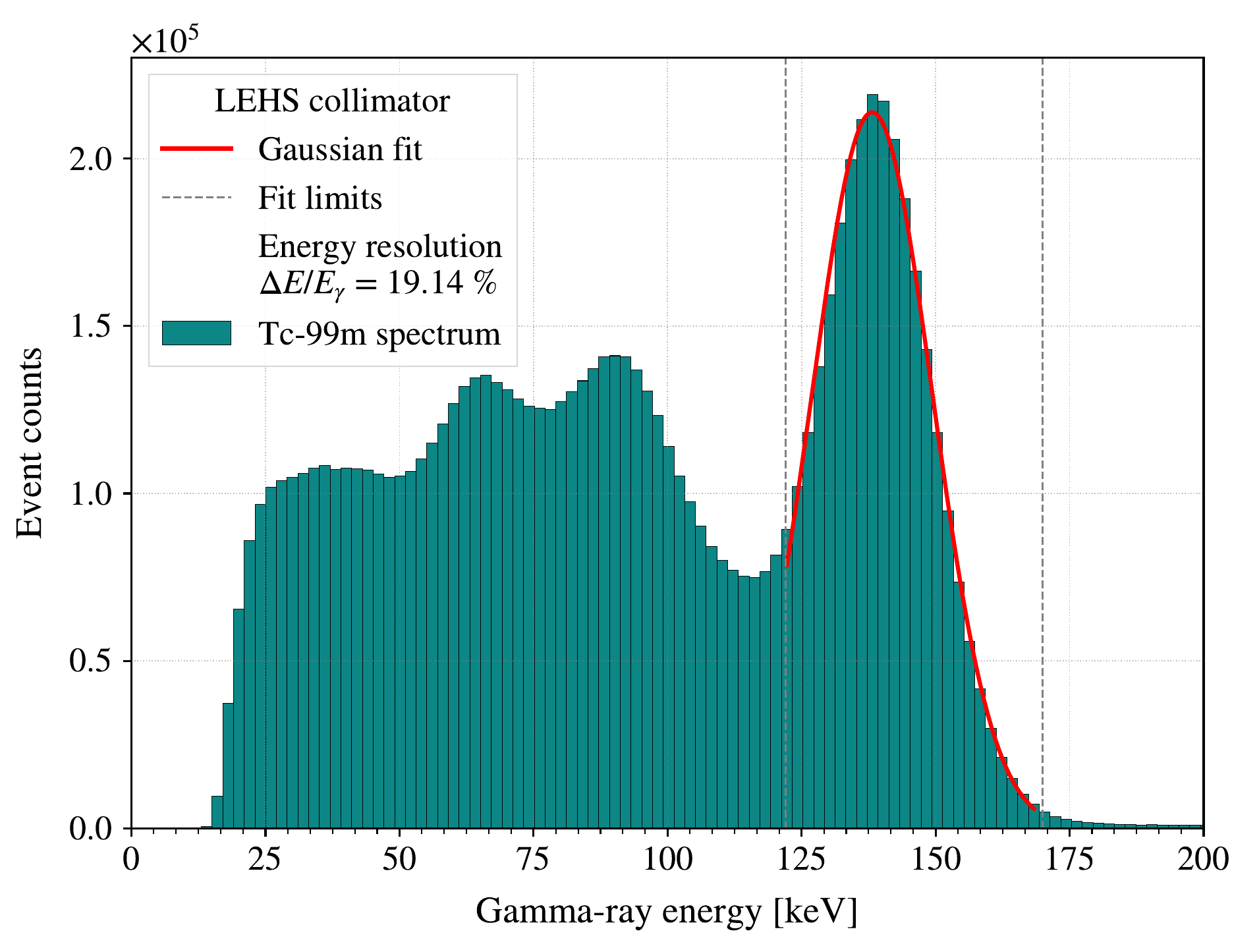}
    \caption{\textbf{Left:} Corrected spectrum obtained with POSiCS (LEHR collimator) over the whole UFOV. \textbf{Right:} Corrected spectrum obtained with POSiCS (LEHR collimator) over the whole UFOV. Compton backscattering and Compton edge peaks are visible, before the photopeak at 140~$\rm{keV}$.}
    \label{fig:energy_resolution}
\end{figure*}

The achieved energy resolution with the LEHS collimator is 19.14\%~$\pm$~0.26\%, which aligns with the typical energy resolutions reported in the literature for SFOV gamma cameras~\cite{Farnworth2023}. However, this result is relatively sharp for a LYSO-based device~\cite{Nakanishi2017, Deprez2014, Morozov2017}. This energy resolution is enough to distinguish the photopeak from the nearby structures in the energy spectrum, namely the backscattering peak (located at about 90 keV, as expected) and the Compton edge: the third peak from the Right in~\autoref{fig:energy_resolution}.
\par With the LEHR collimator, it was possible to measure an energy resolution of 19.94\%~$\pm$~0.20\%, which, as expected, is comparable with the resolution achieved through the LEHS collimator. Energy resolution is primarily driven by the scintillator (the number of photons produced per scintillation event), the optical coupling to the photodetector, and the photodetector efficiency. Therefore, as assessed by these tests, the collimator does not significantly impact the energy resolution. However, we notice a large amount of scattered gamma rays with the LEHR collimator, which we link to the increased amount of high-Z scatterer material (tungsten) present due to the longer collimator.
\subsection{Collimator lateral leakeage}

\par
\begin{figure}
    \centering
    \includegraphics[width=0.49\linewidth]{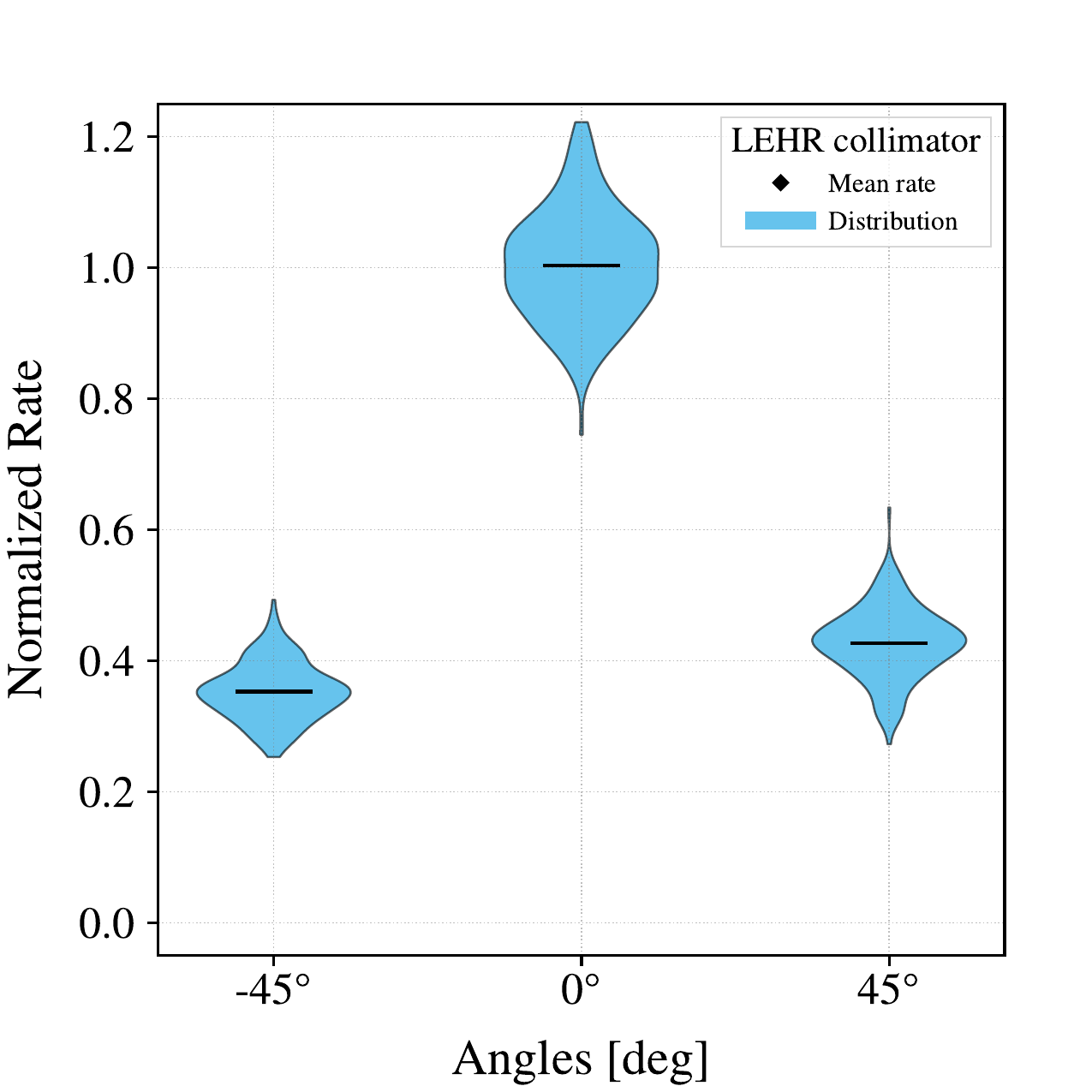}
    \includegraphics[width=0.49\linewidth]{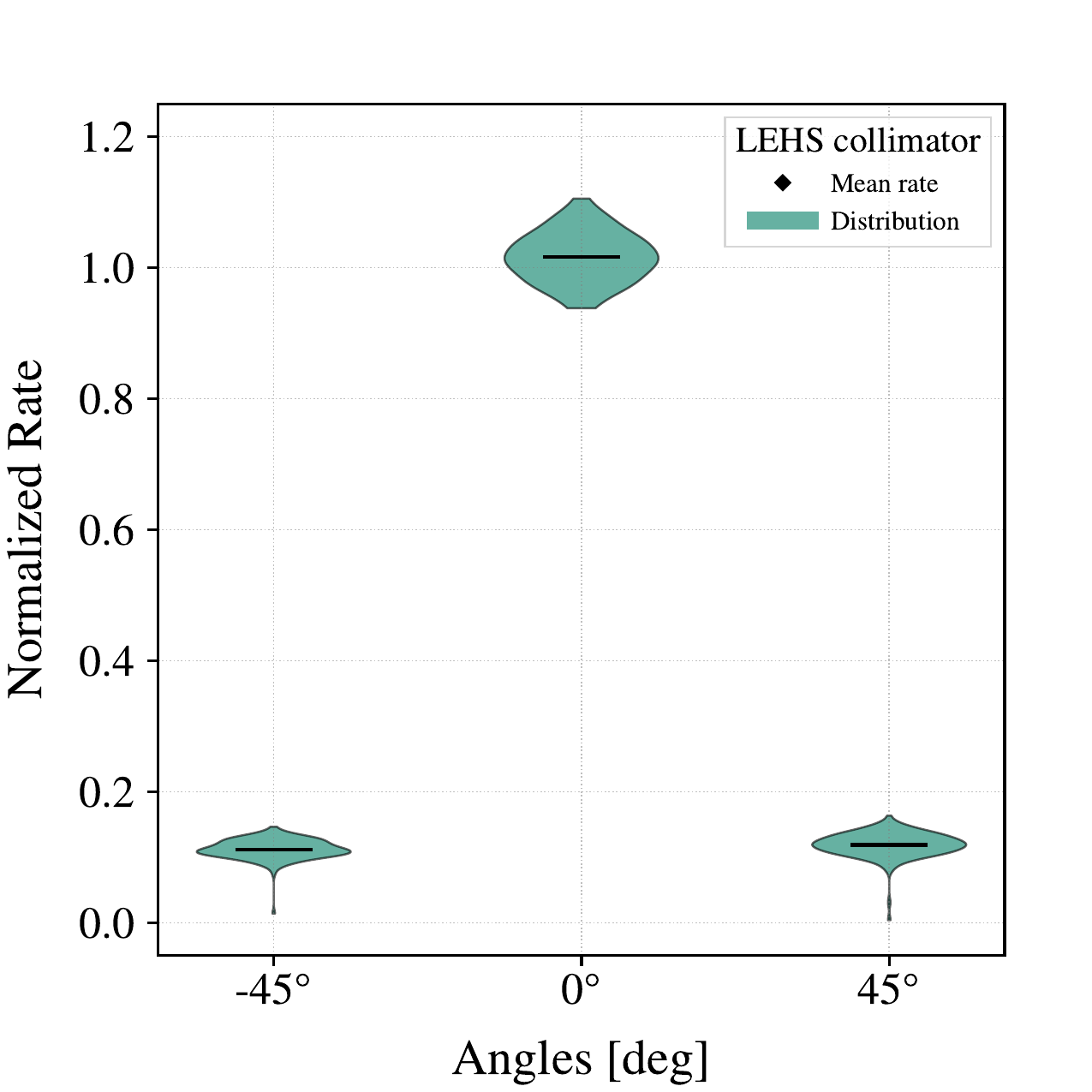}
    \caption{\textbf{Left: } Leakage test results for the LEHR collimator. \textbf{Right: } Leakage test results for the LEHS collimator.}
    \label{fig:septal_leakage_test}
\end{figure}

 The tests with both collimators show that the count rate drops significantly when the source is placed outside of the FOV. Moreover, the response is symmetric when placing the source outside of the FOV to the left and Right of the camera. In the case of the LEHS collimator, the average rate drop outside the FOV is 88\%, indicating that the shielding around the scintillator effectively eliminates most of the incoming gamma rays. The LEHR collimator shows an average relative rate drop of only 61\%. Since both collimators have the same shielding thickness, the observed discrepancy may suggest that the greater amount of tungsten in the LEHR collimator leads to increased gamma-ray scattering. As a result, Compton-scattered gamma rays may still reach the scintillator and be detected. This implies that the energy window cut applied around the photopeak is less effective when using the LEHR collimator. The rate as function of angle is shown in~\autoref{fig:septal_leakage_test}, relative to the the mean rate at $0^\circ$ from the FOV normal direction.

\subsection{Camera's response to other radiotracers (Lu-177)}
With the device under test, it was possible to observe a Lu-177 source emitting at different energies. 
\par
\begin{figure}[h!]
    \centering
    \includegraphics[width=1\linewidth]{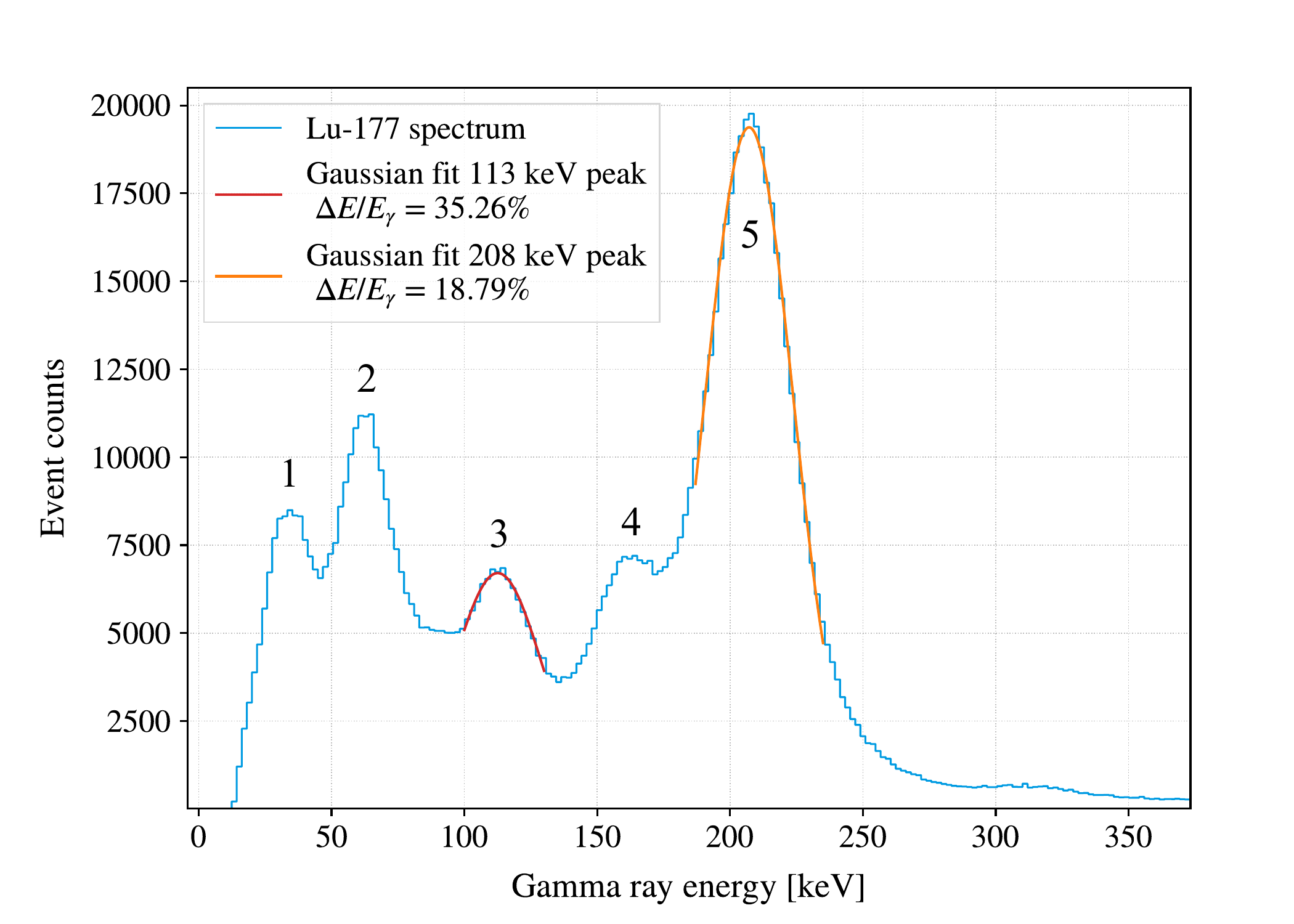}
    \caption{Recorded and rescaled spectrum obtained through the mouse-phantom cavity measurement. The measured energy resolutions at 113~keV and 208~keV are also reported.}
    \label{fig:Lu177_spectrum}
\end{figure}

 The obtained energy spectrum is displayed in~\autoref{fig:Lu177_spectrum}. The main emission peak was found to be at 218~keV, which shows, therefore, a 4.6\% discrepancy with the theoretical value (208~keV). This behaviour is expected, since the camera was calibrated based on a 140.5 keV reference energy, and assumes linear response of the light yield to the reconstructed charge. For this reason, the energy look-up table (LUT) was rescaled to ensure that the main emission peak was centered at 208~keV.

\begin{figure*}[t!]
    \centering
     \includegraphics[width=0.77\linewidth]{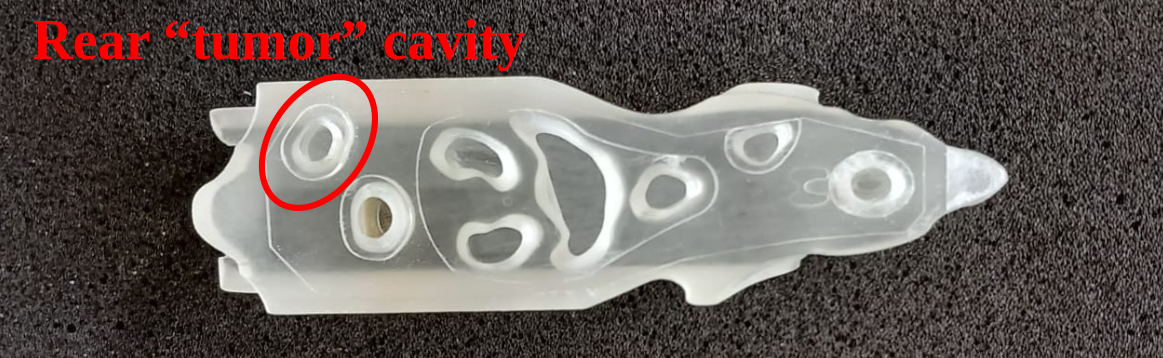}\\
    \vspace{0.5cm}
    \includegraphics[width=0.32\linewidth]{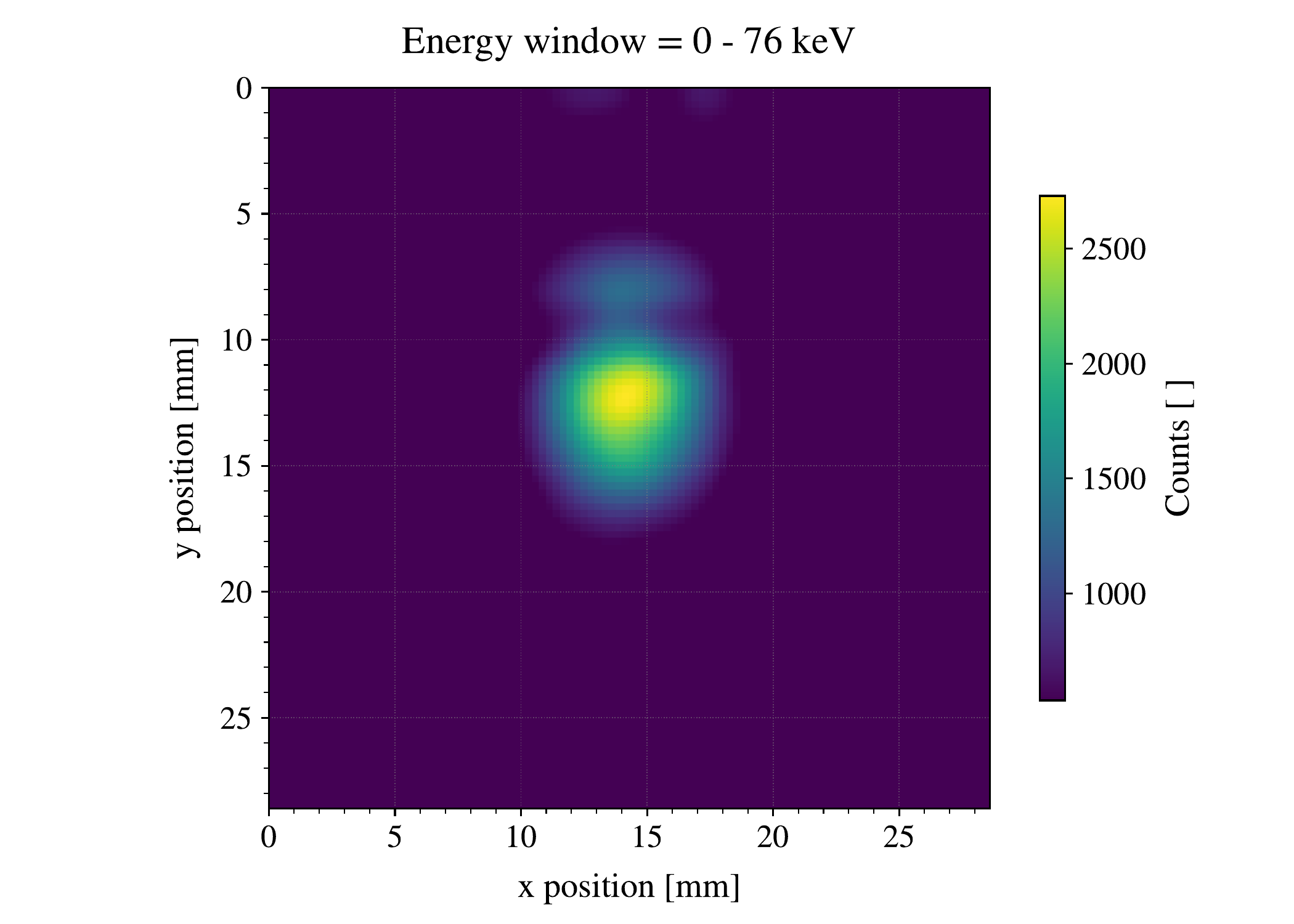}
    \includegraphics[width=0.32\linewidth]{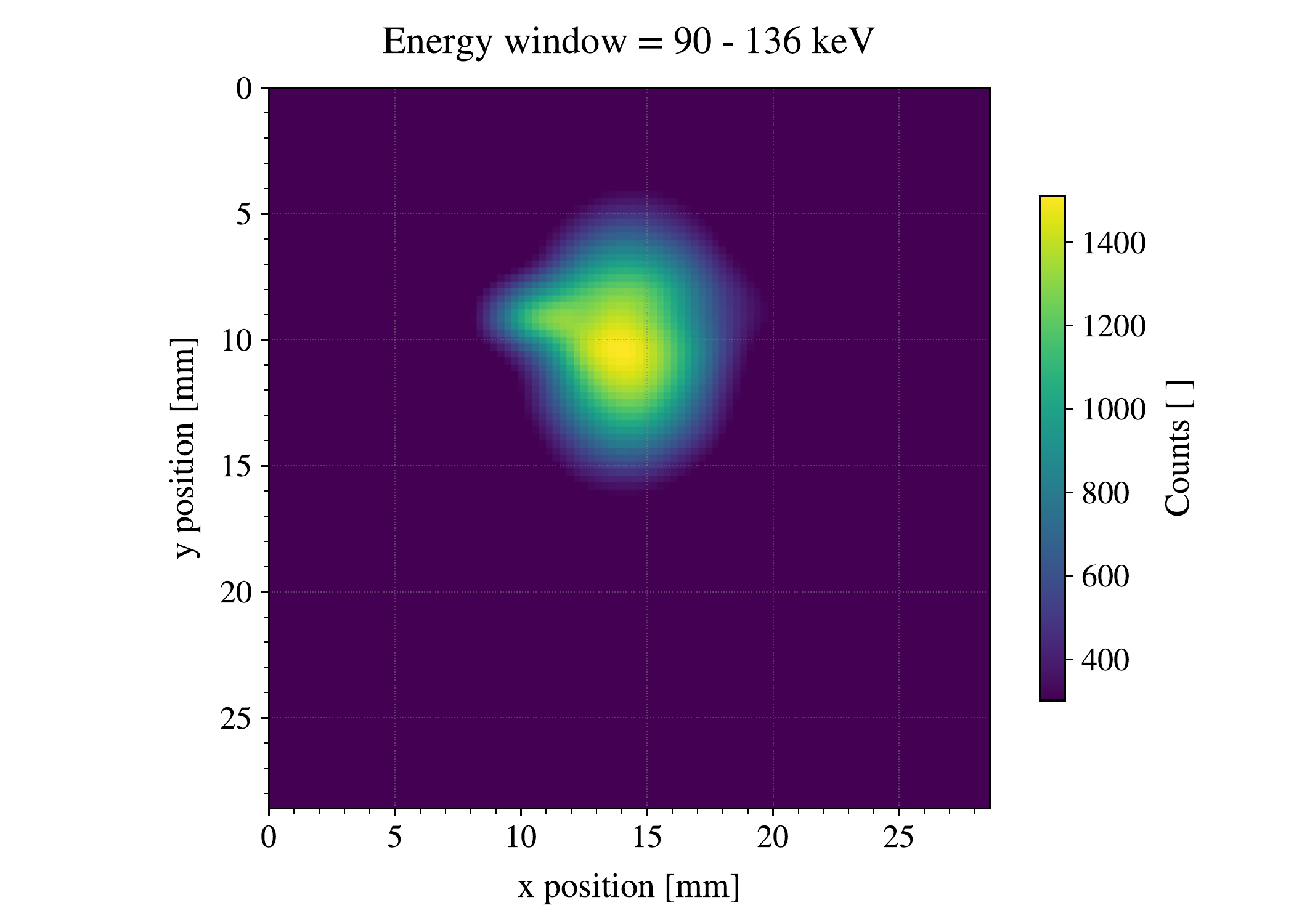}
    \includegraphics[width=0.32\linewidth]{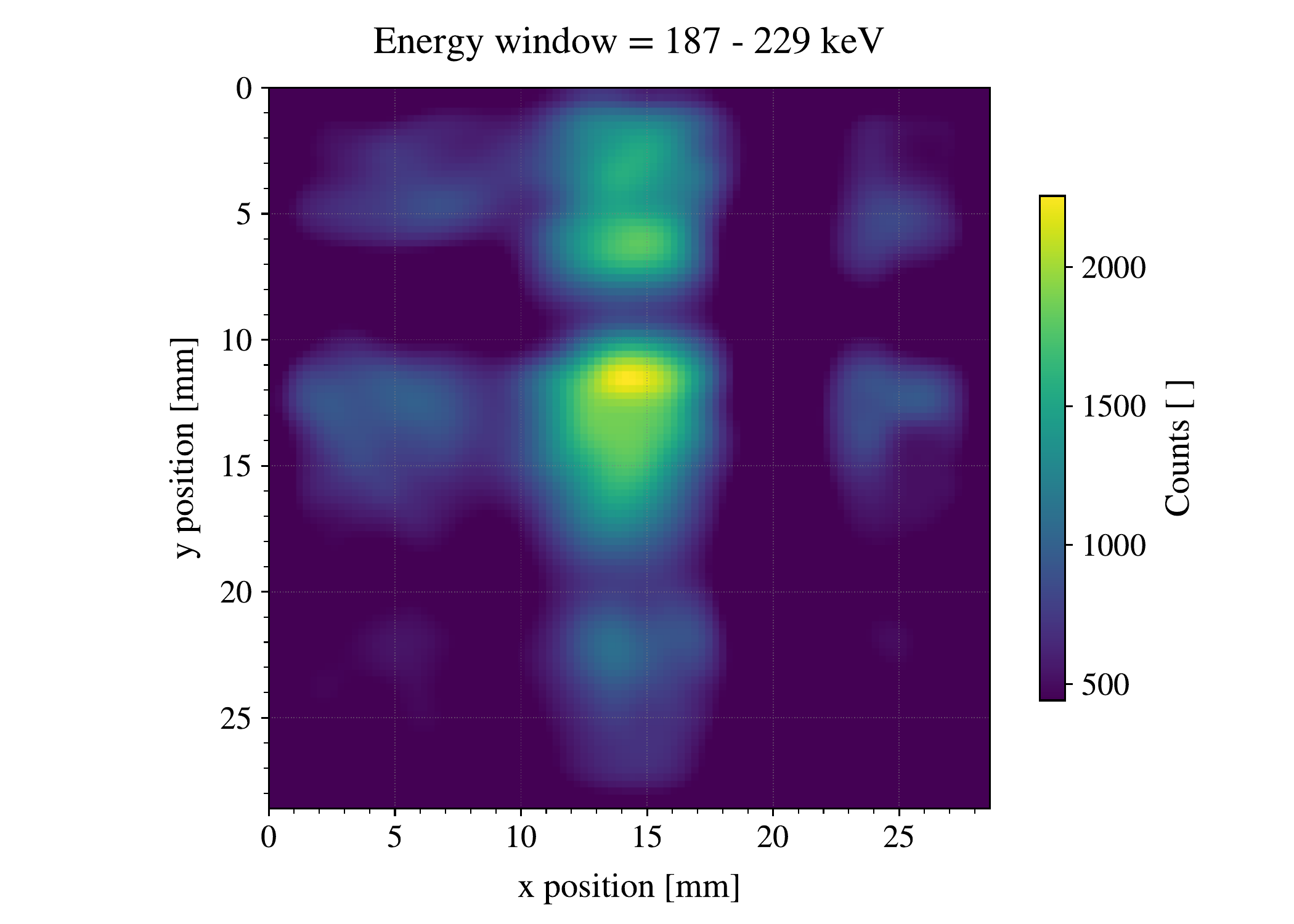}

    \caption{\textbf{Top: }Bioemtech Mice Phantom used for the Lu-177 test. The rear "tumor" cavity was filled with the radiotracer and covered by the phantom's plastic lid. Phantom from~\cite{BiometechMousePhantom}. \textbf{Bottom: }Reconstructed cavity image for three different energy windows after triple correction (Linearity, Energy cut, and uniformity compensation), treated with Gaussian kernel and spline interpolation.}
    \label{fig:Lu177_images}
\end{figure*}

We recognize the first peak as very low-energy X-rays and background radiation, coming from electronics noise and characteristic X-rays from the lead shielding. We identify peak number 2 as the X-ray emissions at 55 keV and 64 keV, which merge into a single peak at 60 keV due to the limited energy resolution of the camera at such low energies, as already observed by~\cite{Sagisaka2024}. Peak number 3 is the 113 keV emission, which is clearly distinguishable and well centered around its reference value. We identify peak number 4 as the Compton shoulder of the main emission line (peak 5). It is also possible to notice a small bump in the spectrum after the 208 keV energy, which could correspond to a weak emission of Lu-177 (321 keV). However, this energy has a low photoelectric absorption probability within the thin scintillator crystal of the POSiCS camera and, therefore, cannot be identified. Moreover, this higher-energy region is also populated by the Lu-176 background due to the presence of this isotope in the LYSO:Ce scintillator (cf. subsection~\ref{ssec:scintillator}).

\par The primary emission energies used for Lu-177 SPECT imaging are the 113 keV and 208 keV gamma-rays. In our case, the energy resolution of these two photopeaks was determined through a Gaussian fit and found to be 35\% and 19\%, respectively. We therefore proceeded with the usual reconstruction pipeline, by defining two 20\% energy windows around these peaks (following the approach of~\cite{Marin2017}) and a large energy window (0-76 keV) to assess the reconstructed images through low-energy X-ray emission. Those peaks were already taken as a reference for imagin with Lu-177 in other works~\cite{Roth2020, Roth2024}.

\par The imaging results after the standard image correction pipeline, Gaussian kernel convolution (following the NEMA standards guidelines~\cite{NEMA2023}), and spline interpolation, are presented in~\autoref{fig:Lu177_images} (more information about the correction pipeline are reported in~\autoref{ssec:corr}). The cavity image can be efficiently reconstructed for both the X-ray and 113 keV energy windows. However, we encountered significant problems in reliably reconstructing an image at 208 keV. As shown in ~\autoref{fig:Lu177_images}, gamma-ray detection in the 208 keV range extends beyond the cavity's region of interest (ROI), with signals detected in pixels outside the area filled with activity. This issue arises from the use of thin collimator septa, which are optimized for lower energies and fail to collimate gamma rays at 208 keV adequately.
\par This test demonstrated that the POSiCS camera, without any modification to its configuration or components, is also suitable for imaging with Lu-177. It effectively detects events associated with the isotope’s X-ray emissions as well as gamma emissions up to its 113 keV peak. However, imaging at higher energies would require the use of a high energy collimator with longer septa. The performance of the current LEHR collimator with Lu-177 is presently under investigation.

\section{Discussion}
\label{ch:5}
\subsection{Comparison with existing literature}
\begin{table*}[t]
\centering
\renewcommand{\arraystretch}{1.3}
\begin{tabular}{lcc}
\hline\hline
\textcolor{blue}{\textbf{POSiCS camera performance}} &
\textcolor{blue}{\textbf{LEHS collimator}} &
\textcolor{blue}{\textbf{LEHR collimator}} \\
\hline\hline

\multicolumn{3}{c}{\textcolor{blue}{\textbf{Spatial resolution}}} \\
\hline
Spatial resolution AEC (122 keV) @0cm [mm]$^*$ & 1.9 $\pm$ 0.1 & 1.4 $\pm$ 0.3 \\
Spatial resolution AEC (122 keV) @2cm [mm]$^*$ & 4.2 $\pm$ 0.1 & 2.4 $\pm$ 0.3 \\
\hline

\multicolumn{3}{c}{\textcolor{blue}{\textbf{Sensitivity}}} \\
\hline
Sensitivity BEC (140.5 keV) @2cm [cps/MBq] & 1444 $\pm$ 31 & 730 $\pm$ 16 \\
Sensitivity AEC (140.5 keV) @2cm [cps/MBq]$^{**}$ & 481 $\pm$ 14 & 134 $\pm$ 8 \\
\hline

\multicolumn{3}{c}{\textcolor{blue}{\textbf{Other parameters}}} \\
\hline
Energy resolution (140.5 keV) @2cm & 19.14\% $\pm$ 0.26\% & 19.94\% $\pm$ 0.20\% \\
Rate drop outside FOV (122 keV) @3cm & 88\% & 61\% \\
Weight [kg] & 0.313 & 0.381 \\
Field Of View (FOV) & $30.8\times30.8~\mathrm{mm}^2$ & $30.8\times30.8~\mathrm{mm}^2$ \\
Useful Field Of View (UFOV) & $28.6\times28.6~\mathrm{mm}^2$ & $28.6\times28.6~\mathrm{mm}^2$ \\
\hline\hline
\end{tabular}

\caption{Performance parameters results presented in this work.
$^*$ Results of resolution measurement with scatterer material.
$^{**}$ Lower limit of the sensitivity for this collimator.
BEC = Before Energy Cut, AEC = After Energy Cut.}
\label{tab:summary}
\end{table*}

In this work, we presented a performance study of the newly developped POSiCS camera.  A summary table of these results is presented in~\autoref{tab:summary}.
\par To reference POSiCS performances within the landscape of already existing IGCs, a comparison with the literature is proposed, starting with the following parameters: Spatial resolution, extrinsic sensitivity, energy resolution and device's weight.

\par
\begin{figure}[h!]
    \centering
    \includegraphics[width=.99\linewidth]{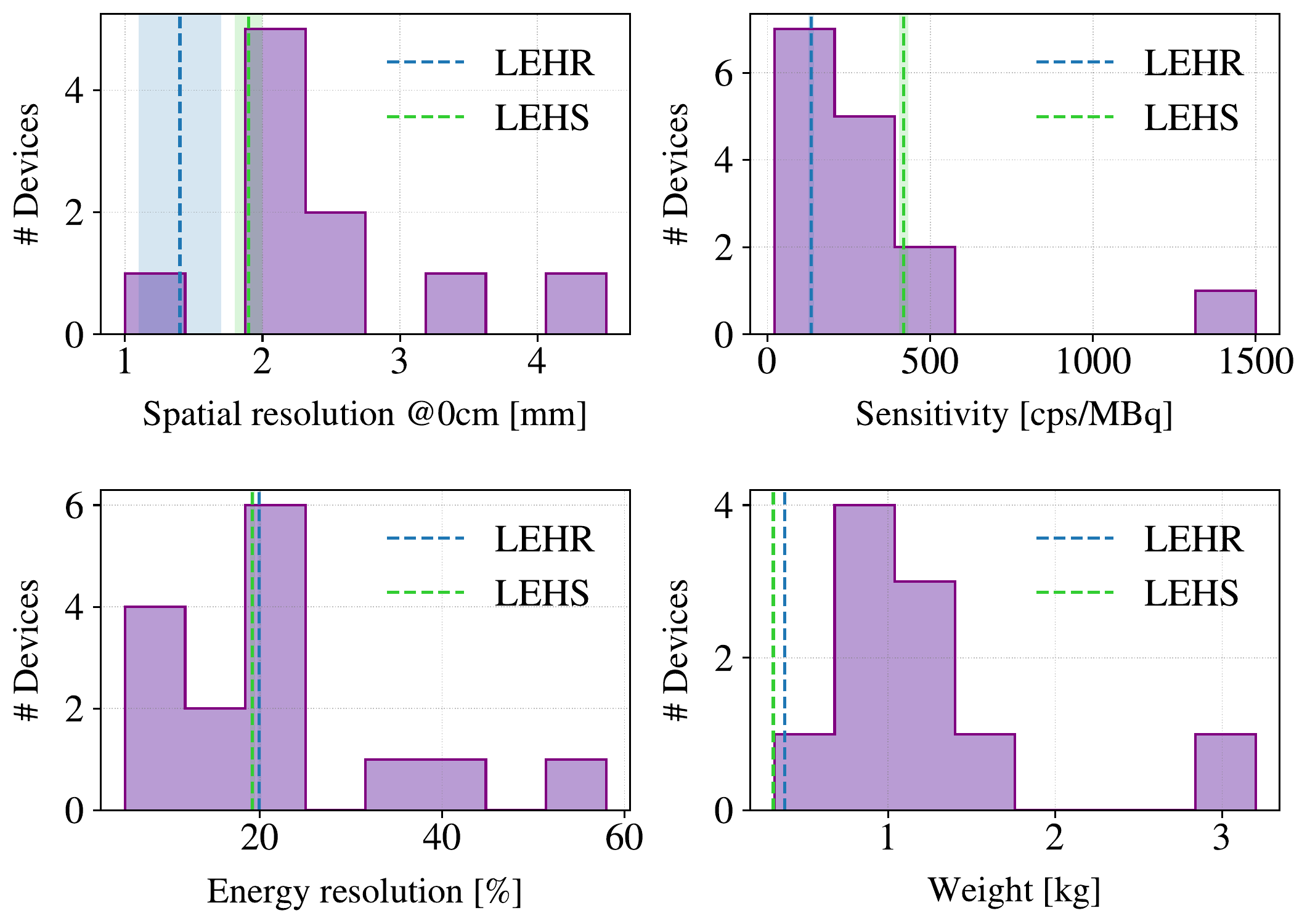}
    \caption{Distribution of the devices performance pareameters repoted in literature, compared to the figure of merits for the POSiCS camera measured in thiw work. The shaded bands represent the experimental uncertainty on POSiCS' parameters.}
    \label{fig:review_comparison}
\end{figure}

 The evaluation of POSiCS performance parameters can be compared with literature using the comprehensive 2023 review on IGCs~\cite{Farnworth2023}. From this review, the relevant performance and physical characteristics of other intraoperative SFOV gamma cameras were extracted. The mean value, the standard deviation, and the minimal and maximal values of these parameters are reported in~\autoref{tab:stats}. Only devices for which the target parameter was available were included in the statistical analysis and comparison. For cameras featuring multiple operating modes (e.g., interchangeable collimators of different lengths), each parameter was selected from the configuration that most favored the corresponding performance metric, to be conservative. The devices parameters distribution compared to the results of this work for the POSiCS camera are presented in~\autoref{fig:review_comparison} and commented below.

\par For spatial resolution, only devices for which the resolution at 0 cm from the source was specified were taken into consideration, in order to be compared with the in-contact performances of POSiCS. It is important to highlight that the evaluation of POSiCS spatial resolution was performed at an energy of 122~keV, while some of the devices used for the comparison evaluated the same parameter at 140~keV. POSiCS shows an extrinsic resolution of a few millimeters even at distances of 4 cm from the target source. This requirement is satisfied with both collimators. The resolution of the camera at 0 cm from the source outperforms 9 devices out of 10 in both LEHR and LEHS modes. However, this trend may vary at longer source-to-camera distances due to the geometrical projection effect of the collimator. Overall, POSiCS meets the millimetric spatial resolution requirement up to a source-to-collimator distance of 4~cm, in agreement with the initial specifications for SLNB and respecting the requirements for SLNB. 

\par Since the extrinsic sensitivity of a gamma camera is less dependent on the distance from the source~\cite{cherry_ch14}, we opted to consider all the devices with known sensitivity estimation, regardles of the reference source-to-collimator distance from the source at which the parameter was estimated. 

From~\autoref{tab:stats}, we notice a large standard deviation of the devices' sensitivity statistics, which is driven by the different purposes of the cameras presented in the 2023 IGC review (some of which target high resolution, and others high sensitivity). The LEHR camera outperforms only 2 out of 15 devices in terms of sensitivity, since the main target of this configuration is to have a high-precision in image reconstruction, and is therefore more suited for higher activity sources. The LEHS mode, instead, shows a sensitvity well above average, outperforming 12 devices out of 15. Only one device has a significantly higher sensitivtiy (about 1500~cps/MBq). This specific gamma camera leverages a peculiar collimator-scintillator configuration enhancing the device's sensitivity, and represents a far outlier in the statistics.

\par The measured energy resolution of the POSiCS camera is in line with the average of 21.4\% presented in the review (even if this value, once again, was evaluated at 122~keV and 140~keV, depending on the device). The cameras showing a significantly sharper energy resolution with respect to POSiCS are either scintillator cameras based on scintillators with higher light yields, such as LaBr$_3$:Ce or IGCs performing imaging with a direct detection system, such as CZT detectors that are therefore more suited for spectroscopy. Nevertheless, the observed energy resolution remains competitive, especially when compared to larger LYSO-based systems operating near 140~keV, which often exhibit energy resolutions exceeding 20\%. Moreover, LYSO offers additional benefits: its high density allows for the development of compact detectors without reducing the devices' sensitivity.

\par A comparison of the cameras' weights was also performed. This parameter is of great importance to introduce gamma cameras within the protocol of SNLB and ROLL standard procedures, since it allows a easier manipulation of the device. POSiCS outperforms 9 devices out of 10 with the LEHR collimator configuration, while it outperforms all the devices in LEHS mode, with a weight well below the average of 1.27~kg. Only one device had a weight comparable to POSiCS (about 320~g), but with a FOV of 1.3$\times$1.3~cm$^2$, implying a very small device and hence a reduced weight.

\subsection{Lateral collimator leakage tests}
\par Shielding tests revealed some residual gamma-ray penetration in both collimator. When outside of the FOV, the camera mounted with the LEHS collimator showed a rate drop of 88\%, while in LEHR configuration, the rate drop was of only 61\%. We explain this due to the increased amount of scattering associated to the longer collimator (as proven by the full FOV energy spectrum of the camera), and to the lower amount of accepted photons by the LEHR collimator. The violin plots in~\autoref{fig:septal_leakage_test} show that the distribution of normalized rates is broader for the HR collimator than for the HS configuration. This larger variability is observed because the HR collimator registers a lower absolute rate, as it is designed to reject a larger fraction of incoming trajectories (i.e. reject more gamma-rays). Since the leakage from the camera's sides is expected to be similar for both collimators (given their identical shielding thickness) the same absolute leakage produces a larger relative contribution in the HR case. As a result, the relative rate reduction is smaller for the HR collimator. This implies that the lateral leakage has a stronger impact on image reconstruction for the LEHR configuration, and might dominate when observing low activity sources within the FOV (for instance when observing near the injection zone). 
\par The analysis shows that an improved collimator design with enhanced lateral shielding could mitigate this penetration effects, and that might be required especially for the HR collimator. 

\subsection{POSiCS performances at different energies}
\par Tests with $^{177}$Lu were performed due to the increased need for radiation monitoring in $^{177}$Lu-DOTA-TATE and $^{177}$Lu-PSMA procedures, and to test POSiCS performances at different energies. 
\par The tests revealed that it was possible to image a liquid $^{177}$Lu source in a plastic mouse phantom, with the two energy windows:
0-76~keV and 90-136~keV, targeting low energy Lu-177 emissions (X-Rays) with the first window and the 113~keV emission line, with the second window. The third emission line, at 208~keV, while composing the larger fraction of detected gamma-rays, was not suited for imaging. The collimator walls of the LEHS collimator were too thin to correctly absorb rejected trajectory, and this results in a strong leakage in the whole camera FOV, as shown in~\autoref{fig:Lu177_images}.

Tests with the LEHR collimator are currently being performed and the possibility of a Lu-177 based collimator with thicker septa is under evaluation. 
\par The higher septal penetration due to the 208~keV emission of Lu-177 is a well known problem in SFOV gamma cameras, and has been reported for other devices~\cite{Roth2020, Roth2024}. Once again, the assestment and evaluation of a Lu-177 dedicated collimator could allow the POSiCS camera to provide imaging at 208~keV with the simple need of swapping the device's head.

\begin{table}[h]
\centering
\begin{tabular}{lccccc}
\toprule
Parameter & \# Devs & Mean & Std & Min & Max \\
\midrule
Spatial resolution [mm] & 10 & 2.44 & 0.95 & 1.0 & 4.5 \\
Sensitivity [cps/MBq] & 15 & 317 & 356 & 21 & 1500 \\
Energy resolution [\%] & 15 & 21.3 & 13.7 & 5.2 & 58 \\
Weight [kg] & 10 & 1.27 & 0.76 & 0.32 & 3.2 \\
\bottomrule
\end{tabular}
\caption{Statistics of intraoperative gamma cameras from literature.}
\label{tab:stats}
\end{table}

\section{Conclusions}
\label{ch:6}
The POSiCS camera was developed as a lightweight, compact, and wireless IGC to assist surgeons during SLNB and ROLL procedures. Integrating rapid, user-friendly imaging systems in surgical settings can reduce procedural invasiveness, thus enhancing both treatment efficacy and postoperative outcomes.

\par POSiCS demonstrated high spatial resolution when equipped with a LEHR collimator and showed good sensitivity in its LEHS collimator configuration. By comparing the results described in this work with current state-of-the-art IGCs, we demonstrated that POSiCS performances are comparable to, or superior to, currently available systems~\cite{Farnworth2023, Tsuchimochi_2013}, while offering additional benefits, such as a lightweight module (less than 400 g) and wireless communication. These features significantly improve usability within the surgical environment. 

\par Furthermore, the camera successfully detected gamma emissions from Lu-177, demonstrating imaging capability up to the 113 keV photon emission line of this isotope. It was also possible to produce reliable images of the source by detecting lower-energy X-rays. These findings suggest that the device’s field of applications extends beyond traditional techniques involving the administration of Tc-99m, and may be suited for applications in theranostics and radiotherapeutic dosimetry, without the need for modifications to the detector system.

\section*{List of abbreviations}

\noindent\begin{tabular}{@{}ll@{}}
AEC & After energy cut \\
ADC & Analog-to-digital converter \\
BEC & Before energy cut \\
CZT & Cadmium–zinc–telluride \\
CoG & Center of gravity \\
DAC & Digital-to-analog converter \\
FOV & Field of view \\
FGS & Fluorescence-guided surgery \\
GATE & Geant4 Application for Tomographic Emission \\
HR & High-resolution (mode or collimator) \\
HS & High-sensitivity (mode or collimator) \\
IGC & Intraoperative gamma camera \\
LEHR & Low-energy high-resolution (collimator) \\
LEHS & Low-energy high-sensitivity (collimator) \\
LG-SiPM & Linearly graded silicon photomultiplier \\
LUT & Look-up table \\
LYSO:Ce & Lutetium–yttrium oxyorthosilicate, cerium-doped \\
NEMA & National Electrical Manufacturers Association \\
PET & Positron emission tomography \\
PMMA & Polymethyl methacrylate \\
PS & Position-sensitive\\
RGS & Radio-guided surgery \\
ROI & Region of interest \\
ROLL & Radioguided occult lesion localization \\
SC & Slow control \\
SPECT & Single-photon emission computed tomography \\
SiPM & Silicon photomultiplier \\
SLNB & Sentinel lymph node biopsy \\
SPAD & Single-photon avalanche diode \\
TIA & Transimpedance amplifier \\
UFOV & Useful field of view \\
\end{tabular}

\section*{Declarations}

\subsection*{Availability of data and material}
The datasets used and/or analysed during the current study are available from the corresponding author on reasonable request.
\subsection*{Competing interests}
The authors declare no competing interests.
 \subsection*{Authors' contributions}
AR wrote the article, performed all tests, implemented the analysis code, and contributed to the image correction algorithm. 
FA provided the detector and wireless module, performed tests on the electronics, and revised the article. 
CA implemented the image reconstruction algorithm, contributed to the conceptualization, performed part of the tests, and revised the article. 
HA performed tests, contributed to the conceptualization, participated in the drafting, and revised the article. 
DdV supervised the entire project, contributed to the conceptualization, contributed to the drafting, and revised the article. 
AG provided resources, contributed to validation, and supported the hardware development. 
HZ supervised the work, contributed to the conceptualization, and revised the article. 
All authors read and approved the final manuscript.

 \subsection*{Aknowledgements}
 \par This study was conducted as part of the POSiCS project. POSiCS-2 was funded for the second phase of the ATTRACT European initiative in June 2022, following the Horizon 2020 call. The Consortium Partners receive funding from the European Union for ATTRACT under Grant Agreement no. 101004462 and the Horizon 2020 Framework Programme for Research and Innovation (2014–2020). 
 \par\noindent The POSiCS project is being carried out jointly by a consortium between  \href{https://www.unige.ch/dpnc/en/groups/domenico-della-volpe/home/}{University of Geneva (UNIGe)} (Department of Nuclear and Corpuscolar Physics, \href{http://www.pinlab.ch/}{PIN lab} at Hôpitaux Universitaires de Genève (HUG), and \href{https://sd.fbk.eu/en/crs/}{Fondazione Bruno Kessler (FBK)} in Italy. The position-sensitive LG-SiPM is based on the patented technology (EP3063559) developed by FBK.

\printbibliography

\end{document}